\documentclass[twocolumn]{aastex631}
\usepackage{graphicx}
\usepackage[flushleft]{threeparttable}
\usepackage{blindtext}
\usepackage{amsmath}
\usepackage{mathtools}
\usepackage{multirow}
\usepackage{comment}

\turnoffeditone

\DeclareRobustCommand{\ion}[2]{%
\relax\ifmmode
\ifx\testbx\f@series
{\mathbf{#1\,\mathsc{#2}}}\else
    {\mathrm{#1\,\mathsc{#2}}}\fi
\else\textup{#1\,{\mdseries\textsc{#2}}}%
\fi}

\begin{document}
\shortauthors{Jung et al.}
\shorttitle{JWST/NIRSpec Observations of $z\simeq7.47-7.75$ LAEs}
\def\nar{New Astron.}
\def\na{New Astron.}
\title{CEERS: Diversity of Lyman-Alpha Emitters during the Epoch of Reionization}

\author[0000-0003-1187-4240]{Intae Jung}
\affil{Space Telescope Science Institute, 3700 San Martin Drive Baltimore, MD 21218, USA}

\correspondingauthor{Intae Jung}
\email{ijung@stsci.edu}

\author[0000-0001-8519-1130]{Steven L. Finkelstein}
\affiliation{Department of Astronomy, The University of Texas at Austin, Austin, TX, USA}

\author[0000-0002-7959-8783]{Pablo Arrabal Haro}
\affiliation{NSF's National Optical-Infrared Astronomy Research Laboratory, 950 N. Cherry Ave., Tucson, AZ 85719, USA}

\author[0000-0001-5414-5131]{Mark Dickinson}
\affiliation{NSF's National Optical-Infrared Astronomy Research Laboratory, 950 N. Cherry Ave., Tucson, AZ 85719, USA}

\author[0000-0001-7113-2738]{Henry C. Ferguson}
\affil{Space Telescope Science Institute, 3700 San Martin Drive Baltimore, MD 21218, USA}

\author[0000-0001-6251-4988]{Taylor A. Hutchison}
\altaffiliation{NASA Postdoctoral Fellow}
\affiliation{Astrophysics Science Division, NASA Goddard Space Flight Center, 8800 Greenbelt Rd, Greenbelt, MD 20771, USA}

\author[0000-0001-9187-3605]{Jeyhan S. Kartaltepe}
\affiliation{Laboratory for Multiwavelength Astrophysics, School of Physics and Astronomy, Rochester Institute of Technology, 84 Lomb Memorial Drive, Rochester, NY 14623, USA}

\author[0000-0003-2366-8858]{Rebecca L. Larson}
\altaffiliation{NSF Graduate Fellow}
\affiliation{Department of Astronomy, The University of Texas at Austin, Austin, TX, USA}

\author[0000-0002-6386-7299]{Raymond C. Simons}
\affiliation{Department of Physics, 196 Auditorium Road, Unit 3046, University of Connecticut, Storrs, CT 06269, USA}

\author[0000-0001-7503-8482]{Casey Papovich}
\affiliation{Department of Physics and Astronomy, Texas A\&M University, College Station, TX, 77843-4242 USA}
\affiliation{George P.\ and Cynthia Woods Mitchell Institute for Fundamental Physics and Astronomy, Texas A\&M University, College Station, TX, 77843-4242 USA}

\author[0000-0002-7464-7857]{Hyunbae Park}
\affiliation{Lawrence Berkeley National Laboratory, CA 94720, USA}
\affiliation{Berkeley Center for Cosmological Physics, UC Berkeley, CA 94720, USA}

\author[0000-0001-8940-6768]{Laura Pentericci}
\affiliation{INAF - Osservatorio Astronomico di Roma, via di Frascati 33, 00078 Monte Porzio Catone, Italy}

\author[0000-0002-1410-0470]{Jonathan R. Trump}
\affiliation{Department of Physics, 196 Auditorium Road, Unit 3046, University of Connecticut, Storrs, CT 06269, USA}

\author[0000-0001-5758-1000]{Ricardo O. Amor\'{i}n}
\affiliation{Instituto de Investigaci\'{o}n Multidisciplinar en Ciencia y Tecnolog\'{i}a, Universidad de La Serena, Raul Bitr\'{a}n 1305, La Serena 2204000, Chile}
\affiliation{Departamento de Astronom\'{i}a, Universidad de La Serena, Av. Juan Cisternas 1200 Norte, La Serena 1720236, Chile}

\author[0000-0001-8534-7502]{Bren E. Backhaus}
\affiliation{Department of Physics, 196 Auditorium Road, Unit 3046, University of Connecticut, Storrs, CT 06269, USA}

\author[0000-0002-0930-6466]{Caitlin M. Casey}
\affiliation{Department of Astronomy, The University of Texas at Austin, Austin, TX, USA}

\author[0000-0001-8551-071X]{Yingjie Cheng}
\affiliation{University of Massachusetts Amherst, 710 North Pleasant Street, Amherst, MA 01003-9305, USA}

\author[0000-0001-7151-009X]{Nikko J. Cleri}
\affiliation{Department of Physics and Astronomy, Texas A\&M University, College Station, TX, 77843-4242 USA}
\affiliation{George P.\ and Cynthia Woods Mitchell Institute for Fundamental Physics and Astronomy, Texas A\&M University, College Station, TX, 77843-4242 USA}

\author[0000-0003-1371-6019]{M. C. Cooper}
\affiliation{Department of Physics \& Astronomy, University of California, Irvine, 4129 Reines Hall, Irvine, CA 92697, USA}

\author[0000-0003-3881-1397]{Olivia R. Cooper}
\altaffiliation{NSF Graduate Fellow}
\affiliation{Department of Astronomy, The University of Texas at Austin, Austin, TX, USA}

\author[0000-0003-2098-9568]{Jonathan P. Gardner}
\affiliation{Astrophysics Science Division, NASA Goddard Space Flight Center, 8800 Greenbelt Rd, Greenbelt, MD 20771, USA}

\author[0000-0003-1530-8713]{Eric Gawiser}
\affiliation{Department of Physics and Astronomy, Rutgers, the State University of New Jersey, Piscataway, NJ 08854, USA}

\author[0000-0002-5688-0663]{Andrea Grazian}
\affil{INAF--Osservatorio Astronomico di Padova, 
Vicolo dell'Osservatorio 5, I-35122, Padova, Italy}

\author[0000-0001-6145-5090]{Nimish P. Hathi}
\affil{Space Telescope Science Institute, 3700 San Martin Drive Baltimore, MD 21218, USA}

\author[0000-0002-3301-3321]{Michaela Hirschmann}
\affiliation{Institute of Physics, Laboratory of Galaxy Evolution, Ecole Polytechnique Fédérale de Lausanne (EPFL), Observatoire de Sauverny, 1290 Versoix, Switzerland}

\author[0000-0002-6610-2048]{Anton M. Koekemoer}
\affil{Space Telescope Science Institute, 3700 San Martin Drive Baltimore, MD 21218, USA}

\author[0000-0003-1581-7825]{Ray A. Lucas}
\affil{Space Telescope Science Institute, 3700 San Martin Drive Baltimore, MD 21218, USA}

\author[0000-0001-5846-4404]{Bahram Mobasher}
\affiliation{Department of Physics and Astronomy, University of California, 900 University Ave, Riverside, CA 92521, USA}

\author[0000-0002-5269-6527]{Swara Ravindranath}
\affil{Space Telescope Science Institute, 3700 San Martin Drive Baltimore, MD 21218, USA}

\author[0000-0002-4772-7878]{Amber N. Straughn}
\affiliation{Astrophysics Science Division, NASA Goddard Space Flight Center, 8800 Greenbelt Rd, Greenbelt, MD 20771, USA}

\author[0000-0003-3466-035X]{{L. Y. Aaron} {Yung}}
\altaffiliation{NASA Postdoctoral Fellow}
\affiliation{Astrophysics Science Division, NASA Goddard Space Flight Center, 8800 Greenbelt Rd, Greenbelt, MD 20771, USA}

\author[0000-0002-6219-5558]{Alexander de la Vega}
\affiliation{Department of Physics and Astronomy, University of California, 900 University Ave, Riverside, CA 92521, USA}

\submitjournal{the Astrophysical Journal}

\begin{abstract}
We analyze rest-frame ultraviolet to optical spectra of three $z\simeq7.47$ -- $7.75$ galaxies whose Ly$\alpha$-emission lines were previously detected with Keck/MOSFIRE observations, using the JWST/NIRSpec observations from the Cosmic Evolution Early Release Science (CEERS) survey. From NIRSpec data, we confirm the systemic redshifts of these Ly$\alpha$ emitters, and emission-line ratio diagnostics indicate these galaxies were highly ionized and metal poor. We investigate Ly$\alpha$ line properties, including the line flux, velocity offset, and spatial extension.  For the one galaxy where we have both NIRSpec and MOSFIRE measurements, we find a significant offset in their flux measurements ($\sim5\times$ greater in MOSFIRE) and a marginal difference in the velocity shifts. The simplest interpretation is that the Ly$\alpha$ emission is extended and not entirely encompassed by the NIRSpec slit. The cross-dispersion profiles in NIRSpec reveal that Ly$\alpha$ in one galaxy is significantly more extended than the non-resonant emission lines. We also compute the expected sizes of ionized bubbles that can be generated by the Ly$\alpha$ sources, discussing viable scenarios for the creation of sizable ionized bubbles ($>$1\,physical Mpc).  The source with the highest-ionization condition is possibly capable of ionizing its own bubble, while the other two do not appear to be capable of ionizing such a large region, requiring additional sources of ionizing photons.  Therefore, the fact that we detect Ly$\alpha$ from these galaxies suggests diverse scenarios on escape of Ly$\alpha$ during the epoch of reionization. High spectral resolution spectra with JWST/NIRSpec will be extremely useful for constraining the physics of patchy reionization.
\end{abstract}

\section{Introduction}
Reionization models predict different scenarios of, for example, \textit{early} vs. \textit{late} reionization \citep[e.g.,][]{Rosdahl2022a, Kannan2022a, Yung2020a, Yung2020b}, questioning the primary sources of ionizing photons that are responsible for reionizing the neutral intergalactic medium (IGM).  An earlier start of reionization is expected when faint and low-mass galaxies with a higher Lyman-continuum (LyC) escape dominate the ionizing photon budget over bright galaxies \citep{Finkelstein2019b}.  Conversely, a relatively delayed process of reionization is predicted when the contributions from faint galaxies ($M_{\text{UV}}$\,$>$\,-$18$) are subdominant to that from brighter systems \citep{Robertson2015a, Naidu2020a}.  Thus, the relative contributions of faint versus bright galaxies to reionization must be imprinted in the temporal and spatial evolution of reionization, and optimal places for probing the sources of ionizing photons are ionized regions in the IGM where we expect to detect Lyman-alpha (Ly$\alpha$) from galaxies \citep[e.g.,][]{Iliev2006a, Mesinger2008a, Rosdahl2018a, Ocvirk2020a, Smith2021a}.

Ly$\alpha$ observations have been broadly used to trace the existence of neutral gas in the IGM \citep[e.g.,][]{Rhoads2001a, Stark2011a, Pentericci2011a, Tilvi2014a} as the visibility of Ly$\alpha$ is sensitive to the amount of neutral hydrogen in the IGM due to resonant scattering \citep[e.g.,][]{Miralda-Escude1998a, Dijkstra2014a}.  Although Ly$\alpha$ is heavily suppressed by the neutral IGM into the epoch of reionization, largely-ionized bubbles in the IGM should provide channels for escape of Ly$\alpha$ \citep[e.g.,][]{Mason2020a, Park2021a, Qin2021a, Smith2021a}.  Thus, this allows us to locate ionized regions in the mostly neutral IGM by finding Ly$\alpha$-emission lines from reionization-era galaxies. Also, the IGM transmission and line profiles of Ly$\alpha$ to constrain the IGM neutral fraction \citep[e.g.,][]{Mason2018a, Mason2019a, Hoag2019a, Jung2020a, Bolan2022a, Hayes2023a}.

Over the last decade, spectroscopic searches for Ly$\alpha$  from reionization-era galaxies at $z\sim$ 7 -- 9 have been discovering Ly$\alpha$-emission lines particularly from UV-brighter ($M_\text{UV}$\,$\lesssim$\,$-20$) galaxies \citep[e.g.,][]{Finkelstein2013a, Zitrin2015a, Oesch2015a, Roberts-Borsani2016a, Castellano2018a, Hu2021a, Endsley2022a, Larson2022a, Jung2022a, Jung2022b}, compared to rarer findings from fainter ones \citep{Hoag2019a, Roberts-Borsani2022a}.  The higher detection rates for Ly$\alpha$ in brighter galaxies hint that reionization proceeded earlier in galaxy-overdense regions where the bright galaxies typically reside. In hierarchical models, higher-density peaks collapse earlier, and more massive galaxies form at these peaks.  This results in ionized bubbles in the mostly neutral IGM forming first around the brighter reionization-era galaxies \citep{Mesinger2011a, Ocvirk2020a, Kannan2022a}.

While the detection of Ly$\alpha$ from galaxies at the reionization epoch suggests the existence of sizable ionized structures ($\gtrsim$1~physical Mpc; pMpc) in the IGM \citep{Mason2020a, Park2021a, Qin2021a, Smith2021a}, this requires that there be a sufficient supply of ionizing (or LyC) photons within the ionized regions.  To draw conclusions about the main driver of reionization, the contribution of galaxies to the required ionizing emissivity must be probed in detail \citep[e.g.,][]{Finkelstein2019a, Yung2020a, Naidu2020a, Yeh2023a}.  However, the actual supply of LyC photons from galaxies cannot be directly measured from reionization-era galaxies due to IGM attenuation. Instead, it needs to be indirectly inferred from the combination of the LyC escape fraction ($f_{\text{esc}}$) and the ionizing photon production efficiency ($\xi_{\text{ion}}$). 

In this work, we use the JWST/NIRSpec observations from the Cosmic Evolution Early Release Science (CEERS) survey to analyze rest-frame ultraviolet to optical spectra of three $z\simeq7.47$ -- $7.75$ Ly$\alpha$ emitters (LAEs).
We focus on understanding the detailed interstellar medium (ISM) properties based on our emission-line analysis as well as the escape of Ly$\alpha$ with the estimates of the ionized bubble sizes.  With a suite of emission lines detected in the NIRSpec observations of our targets, we measure the ionizing photon production efficiency ($\xi_{\text{ion}}$) and infer the LyC escape fraction ($f_{\text{esc}}$) based on indirect indicators of $f_{\text{esc}}$ \citep[][]{Izotov2018a,Chisholm2018a,Chisholm2022a,Flury2022b,Mascia2023a}.  We then predict the growth of self-driven ionized bubbles around these LAEs to explore possible scenarios for creating sizable ionized bubbles that provide the channels for escape of Ly$\alpha$.

This paper is structured as follows. In Section 2, we describe our spectroscopic targets, NIRSpec and MOSFIRE observations, and data reduction.  We present the emission-line analysis including Ly$\alpha$ emission, giving the measured physical properties of these emission lines as well as the line-ratio diagnostics in Section 3. Section 4 discusses the ISM properties and escape of Ly$\alpha$ photons. We then summarize our conclusions in Section 5.  In this work, we assume the Planck cosmology \citep{Planck-Collaboration2016a} with $H_0$ = 67.8\,km\,s$^{-1}$\,Mpc$^{-1}$, $\Omega_{\text{M}}$ = 0.308, and $\Omega_{\Lambda}$ = 0.692. The Hubble Space Telescope (HST) F606W, F814W, F105W, F125W, F140W, and F160W bands are referred to as $V_{606}$, $I_{814}$, $Y_{105}$, $J_{125}$, $JH_{140}$ and $H_{160}$, respectively.  All magnitudes in this work are quoted in the AB system \citep{Oke1983a}, and all errors mentioned in this paper represent 1$\sigma$ uncertainties (or central 68\% confidence ranges) unless stated otherwise.

\section{Observations and Data Reduction}
The JWST/NIRSpec \citep{Jakobsen2022a, Boker2023a} observations of our targets were obtained as part of the NIRSpec observations in the Cosmic Evolution Early Release Science (CEERS; ERS 1345, PI: S. Finkelstein) Survey in the CANDELS \citep{Grogin2011a, Koekemoer2011a} Extended Groth Strip (EGS) field.  The CEERS survey will be fully described in S. Finkelstein et al.\ (in preparation, see also \citealt{Finkelstein2022b, Finkelstein2022c}), with the NIRSpec data described in P. Arrabal Haro et al. (in prepapration).

\begin{deluxetable*}{ccccccc}
\tablecaption{Summary of NIRSpec Targets$^\dagger$ \label{tab:targets}} 
\tablehead{
\colhead{Source ID (MPT ID)} & \colhead{R.A. (J2000.0)} &\colhead{Decl. (J2000.0)} & \colhead{$z_{\text{sys}}$} & \colhead{$J_{\text{125}}$}  & \colhead{Grating} & \colhead{Ref.}\\
 \colhead{}  				& \colhead{(degree)}  		& \colhead{(degree)}  	& \colhead{}  	&  \colhead{(mag)} 		& \colhead{} &		 \colhead{} }
\startdata
{z8\_13573 (686)} & {215.15088} & {52.98957} & {7.7528} & {26.5} & {PRISM} & {[1]}\\
{z8\_32350 (689)} & {214.99903} & {52.94197} & {7.5457} & {25.3} & {MR} & {[1]}\\
{z8\_69492 (698)$^{*}$} & {215.05033} & {53.00745} & {7.4710} & {25.2} & {MR} & {[2],[3]}\\
\enddata
\tablecomments{\footnotesize $z_{\rm sys}$: the mean value of emission-line redshifts from H$\beta$ and [\ion{O}{iii}]. References: [1] \cite{Jung2022b}, [2] \cite{Roberts-Borsani2016a}, [3] \cite{Stark2017a}}
\tablenotetext{}{
$^{\dagger}$\footnotesize These objects were revisited in the 2021A MOSFIRE $Y$-band observations \citep{Jung2022b}, and in this work we analyze the MOSFIRE spectra obtained from the 2021A observations.\\
$^{*}$\footnotesize Known as EGS-zs8-2 in \cite{Roberts-Borsani2016a} and \cite{Stark2017a}.
}
\end{deluxetable*}

\subsection{Targets: Ly$\alpha$-Emitters at $z\simeq7.47$\,--\,$7.75$}
The target galaxies discussed in this work were selected from 13 spectroscopically-confirmed $7<z<8$ galaxies in EGS \citep{Oesch2015a, Roberts-Borsani2016a, Tilvi2020a, Jung2022b}, from which Ly$\alpha$-emission lines were detected in previous ground-based observations with the MOSFIRE spectrograph \citep{McLean2012a} on the Keck telescopes.  In the CEERS NIRSpec multi-object spectroscopy (MOS) observations, the NIRSpec's Micro-Shutter Array \citep[MSA;][]{Ferruit2022a} configurations were created to maximize the number of targets from the full samples from various science cases. This resulted in three $z\sim7$ LAEs being targeted in the NIRSpec observations, which includes two UV-luminous galaxies with $M_{\text{UV}}\sim-22$.  The target properties are summarized in Table \ref{tab:targets}. z8\_13573 and z8\_32350 were introduced in \cite{Jung2022b}, and Ly$\alpha$ emission of z8\_69492 (known as EGS-zs8-2) was reported first in \cite{Roberts-Borsani2016a} and \cite{Stark2017a}. 

\begin{figure*}[th]
\centering
\includegraphics[width=1.0\textwidth]{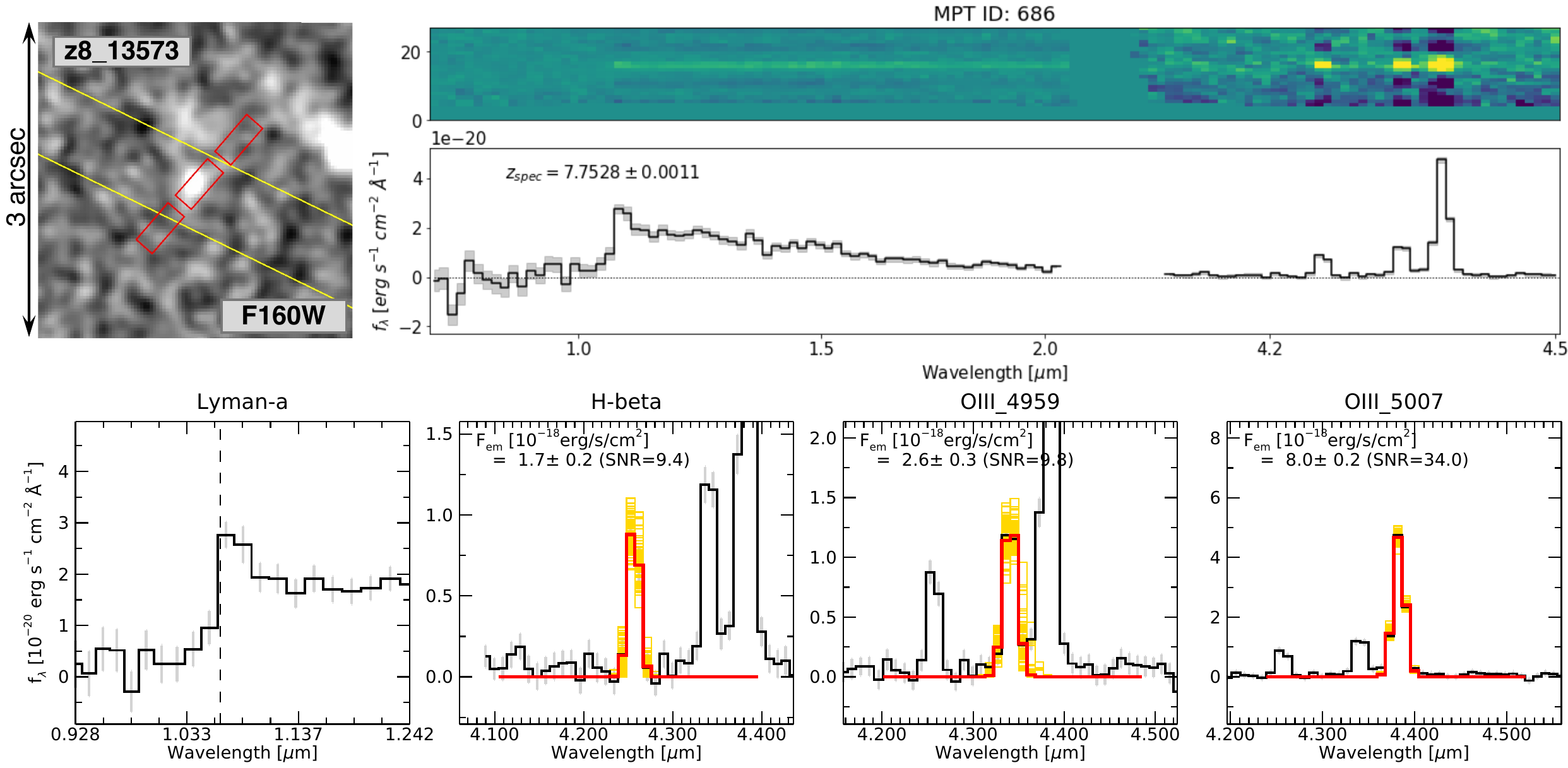}
\caption{(Top left) the NIRSpec MSA shutter configuration on the HST $H_{\text{160}}$ cutout. The three red rectangles show the $0\farcs2\times0\farcs46$ slitlets in NIRSpec observations, and the yellow lines represent the MOSFIRE slit position. (Top right) NIRSpec PRISM 2D and 1D spectra of z8\_13573 which trace the continuum in the short wavelength with Lyman-$\alpha$ break and feature the bright H$\beta$ and [\ion{O}{iii}] lines at their red ends. (Bottom) Zoom-in spectra of individual emission lines seen in the spectra. The black histograms show 1D signals. The best-fit line profiles are shown in red, and the 100 random draws of Monte-Carlo realizations are displayed as yellow. In the Ly$\alpha$ panel, the vertical dashed line represents the systemic wavelength of Ly$\alpha$.} 
\label{fig:686}
\end{figure*}

\begin{figure*}[th]
\centering
\includegraphics[width=1.0\textwidth]{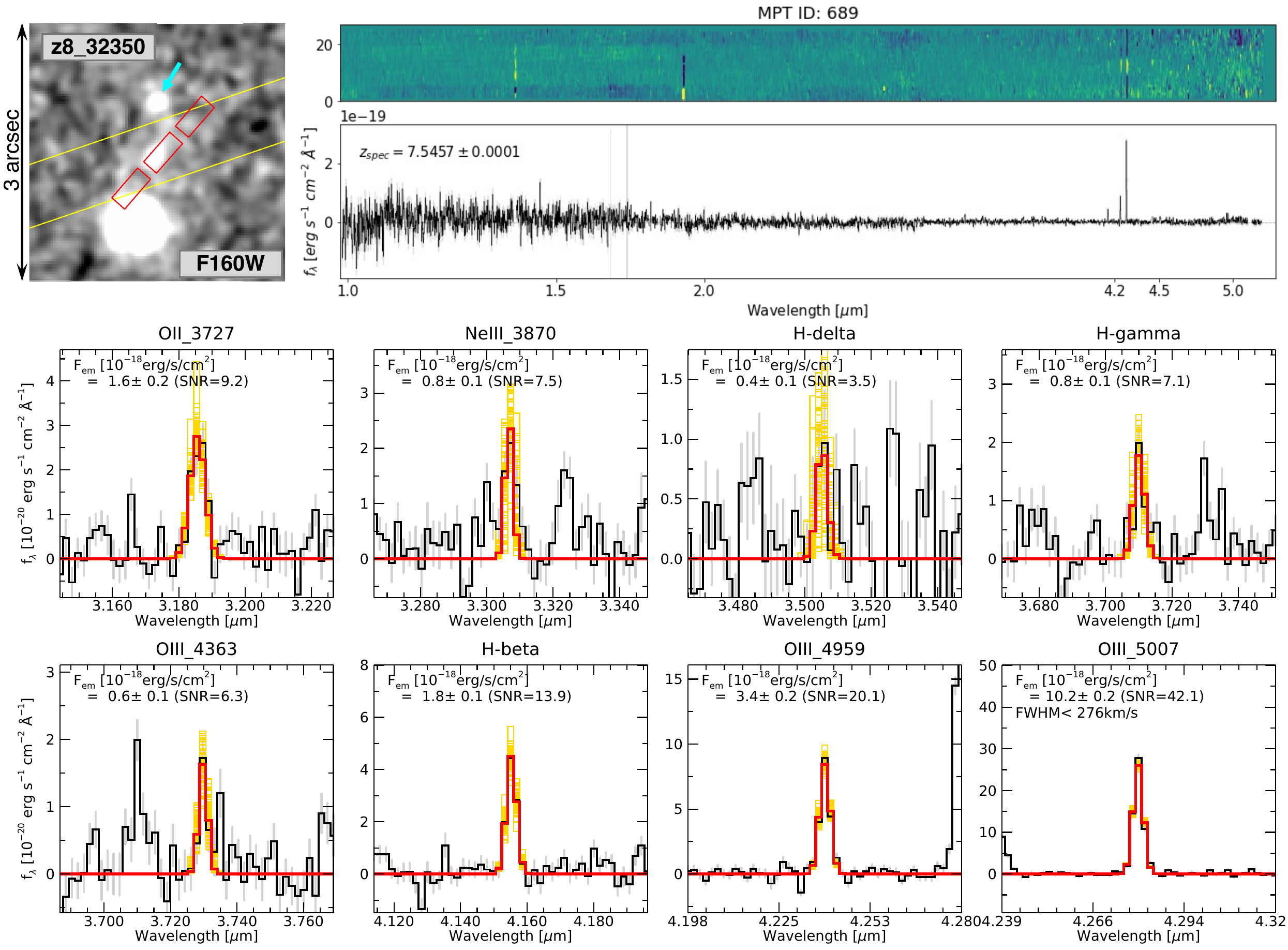}
\caption{Same as Figure \ref{fig:686} but showing the MR composite spectra of G140M/F100LP, G235M/F170LP, and G395M/F290LP of z8\_32350. In the HST cutout, the nearby high-redshift candidate companion is marked with the cyan arrow.} 
\label{fig:689}
\end{figure*}

\begin{figure*}[th]
\centering
\includegraphics[width=1.0\textwidth]{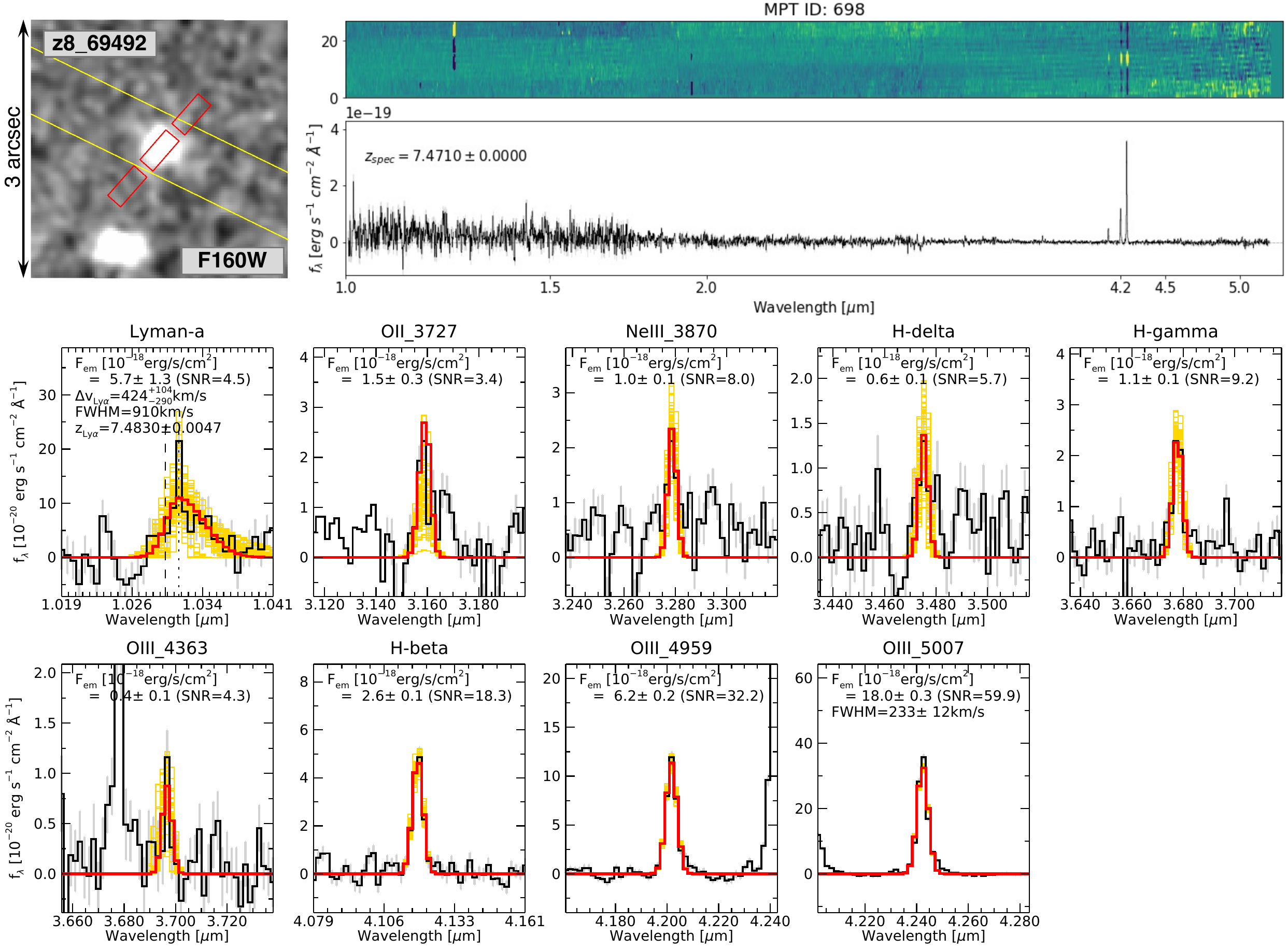}
\caption{Same as Figure \ref{fig:686} but showing the MR composite spectra of G140M/F100LP, G235M/F170LP, and G395M/F290LP of z8\_69492.} 
\label{fig:698}
\end{figure*}

\subsection{JWST/NIRSpec Data}
The NIRSpec observations in the CEERS program include 6 NIRSpec MSA pointings observed with the PRISM ($R\sim30$\,--\,$300$) and CLEAR filter, covering the wavelength range 0.6--5.3$\mu$m, and 6 pointings observed with the medium-resolution (MR; $R\sim1000$) grating/filter pairs  G140M/F100LP, G235M/F170LP, and G395M/F290LP, covering the observed wavelength range of $\sim1$\,--\,$5\mu$m. Four fields have both MR nad PRISM observations.  The NIRSpec targets were observed using three micro shutter slitlets, and a three-point nodding pattern is adopted for background subtraction. Each nod consists of a single integration of 14 groups with 1036\,s exposure, resulting in 3107\,s of total exposure time.  In the top left panels of Figure \ref{fig:686} -- \ref{fig:698}, we show the three slitlets (red rectangles) on the HST $H_{\text{160}}$ images of individual sources.

Here we briefly summarize the main steps of the CEERS NIRSpec data reduction. A detailed description of the data reduction is found in P. Arrabal Haro et al.\ (in preparation).  

The CEERS NIRSpec observations were reduced using the JWST Science Calibration Pipeline version 1.8.5\footnote{https://jwst-pipeline.readthedocs.io/en/latest/} with the JWST Calibration Reference Data System (CRDS) mapping 1027.  We use the \texttt{calwebb\_detector1} pipeline to reduce the Level 0 uncalibrated data products of ``\_uncal.fits'' files. This step includes the saturation check, the bias and dark current subtraction, the correction of the ``snowballs" events caused by cosmic rays, and the 1/f noise correction, generating the count-rate maps (CRMs) from the uncalibrated data.  For the \texttt{jump} step at the detector-level calibration, we used custom parameters for an improved ``snowballs" correction.

The reduced CRMs were then passed through the \texttt{calwebb\_spec2} pipeline. This creates two-dimensional (2D) cutouts of individual slitlets, that are flat-fielded, rectified, and flux-calibrated with the wavelength solution based on the JWST Calibration Reference Data System (CRDS). Then the final 2D spectra were obtained by combining the three nods spectra with the \texttt{calwebb\_spec3} pipeline, and the one-dimensional (1D) spectra were extracted using customized apertures on the 2D spectra to maximize the signal-to-noise ratios (SNRs).  Lastly, for MR-grating-observed targets, a single spectrum was generated per each target, combining all three MR gratings.  Slit-loss corrections were applied with the path-loss correction step in the pipeline, which assumes that the targets are point sources. The resulting 1D and 2D spectra of our targets are shown in the top-right panels in Figures \ref{fig:686} -- \ref{fig:698}.

\subsection{Keck/MOSFIRE data}
Ly$\alpha$-emission lines are detected in the NIRSpec observations from only two of the sources (z8\_13573 with prism and z8\_69492 with MR grating), and the low-resolution prism observations for z8\_13573 precludes detailed analysis of its Ly$\alpha$ line. Fortunately, however, our high-spectral resolution ($R$\,$\sim$\,$3500$) MOSFIRE spectra enable key measurements of the Ly$\alpha$ lines for the sources.

MOSFIRE spectra of our sample were obtained as part of the Keck MOSFIRE $Y$-band observations \citep{Jung2022b} awarded through the NASA allocation (PI: I. Jung). The observations were carried out in two nights in April 2021. The entire observing program consists of four slitmask configurations with a $\sim$3.5\,hr integration time per mask.  z8\_69492 and z8\_13573 were targeted in the first slitmask configuration (EGS\_Y\_2021A\_1) with the moderately good seeing condition of 0.7\arcsec, and z8\_32350 was observed in the slitmask (EGS\_Y\_2021A\_3) with 1.2\arcsec seeing. More details on the observations are found in Table 1 in \cite{Jung2022b}. 

\section{Analysis}
\subsection{Emission-Line Fitting in JWST/NIRSpec data}
To find emission lines, we perform emission-line fitting with the reduced 1D NIRSpec spectra. We first locate strong H$\beta$ and [\ion{O}{iii}] emission lines with the initial guesses based on the spectroscopic redshifts measured from Ly$\alpha$. By performing Gaussian fitting, we measure the fiducial values of systemic redshifts and emission-line widths from the mean values of these lines. Then, we fix the systemic redshifts and the line widths allowed in the Gaussian fitting for all detected emission lines. For the error estimates, we perform Gaussian fitting with 1000 Monte Carlo realizations that are 1D spectra randomly perturbed with corresponding error spectra.  [\ion{O}{ii}] doublets are mostly blended in the MR-grating spectra, but we fit a double Gaussian functional form to [\ion{O}{ii}] doublets to properly model the blended feature. For Ly$\alpha$ emission, we perform asymmetric Gaussian fitting, and the FWHM of Ly$\alpha$ is measured as FWHM$_{\text{Ly}\alpha}$=(FWHM$_{\text{blue}}$+FWHM$_{\text{red}})/2$.  The line fluxes are measured as the area under the best-fit Gaussian, and we subtract the instrumental broadening from the fiducial line width of H$\beta$ and [\ion{O}{iii}].  
We find Ly$\alpha$ emission from two of our LAEs, z8\_13573 and z8\_69492, in NIRSpec observations. We are unable to locate Ly$\alpha$ from z8\_32350 in its NIRSpec G140M grating spectrum. This is probably due to its fainter Ly$\alpha$, compared to that of z8\_69492, below the detection limit. In addition to the prominent detection of H$\beta$ and the [\ion{O}{iii}] doublet from the three sources, we also find additional nebular emission lines from MR grating spectra of the two sources of z8\_32350 and z8\_69492. In the bottom panels of Figures \ref{fig:686} -- \ref{fig:698} we present the individual emission lines of our interest in this work with their flux measurements. Also, the common diagnostics of emission-line ratios are listed in Table \ref{tab:elines}. We will discuss the Ly$\alpha$ properties as well as the ISM properties with the analyses of the line ratios in Section 4.

\subsection{Ly$\alpha$ Measurements from MOSFIRE}
The MOSFIRE observations provide high spectral resolution spectra ($R\sim3500$) of Ly$\alpha$ from our sources. Taking advantage of the high-resolution MOSFIRE spectra, we investigate the detailed properties of the Ly$\alpha$ lines, including Ly$\alpha$ velocity offsets. 

\begin{figure}[t]
\centering
\includegraphics[width=0.9\columnwidth]{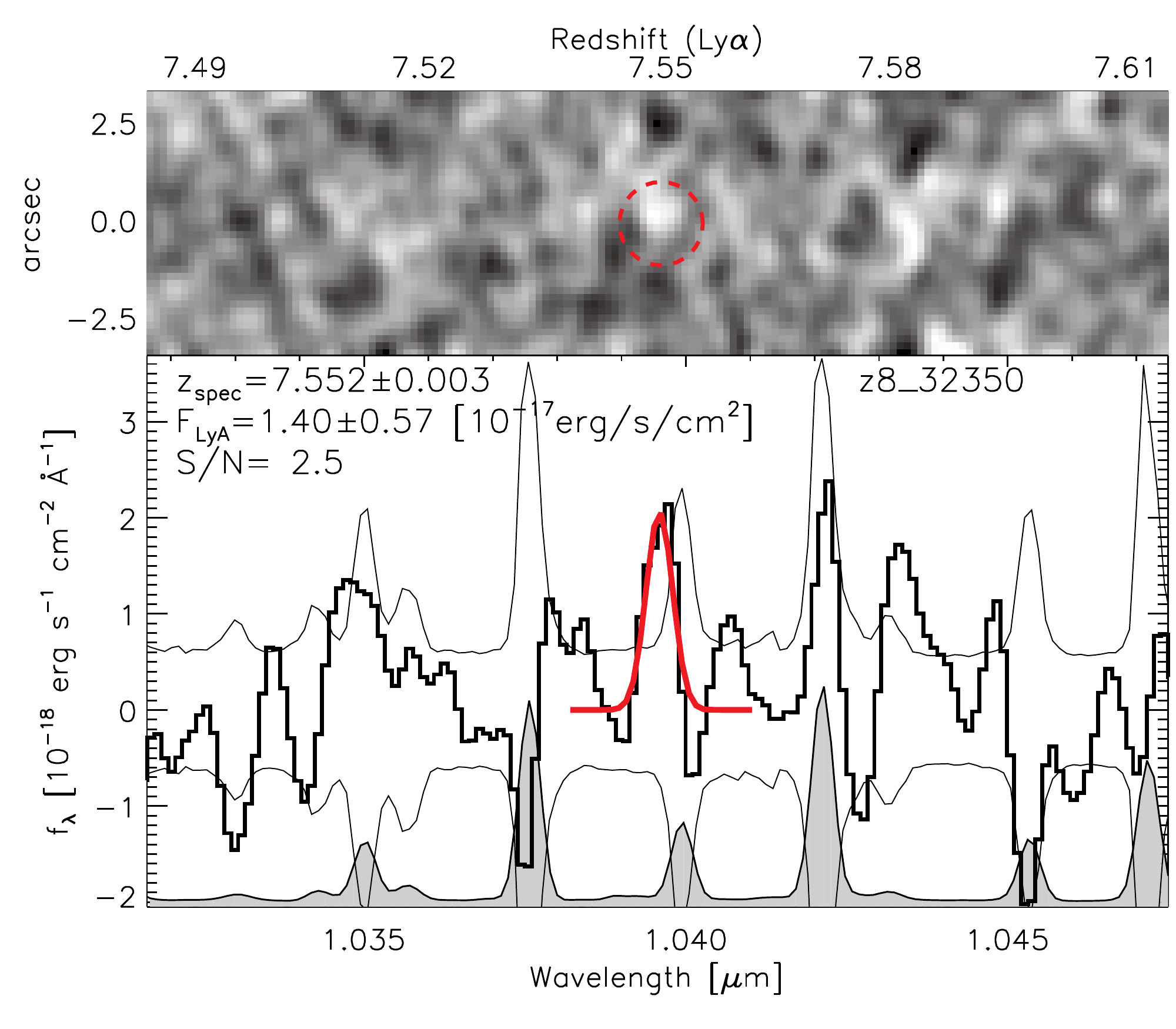}
\includegraphics[width=0.9\columnwidth]{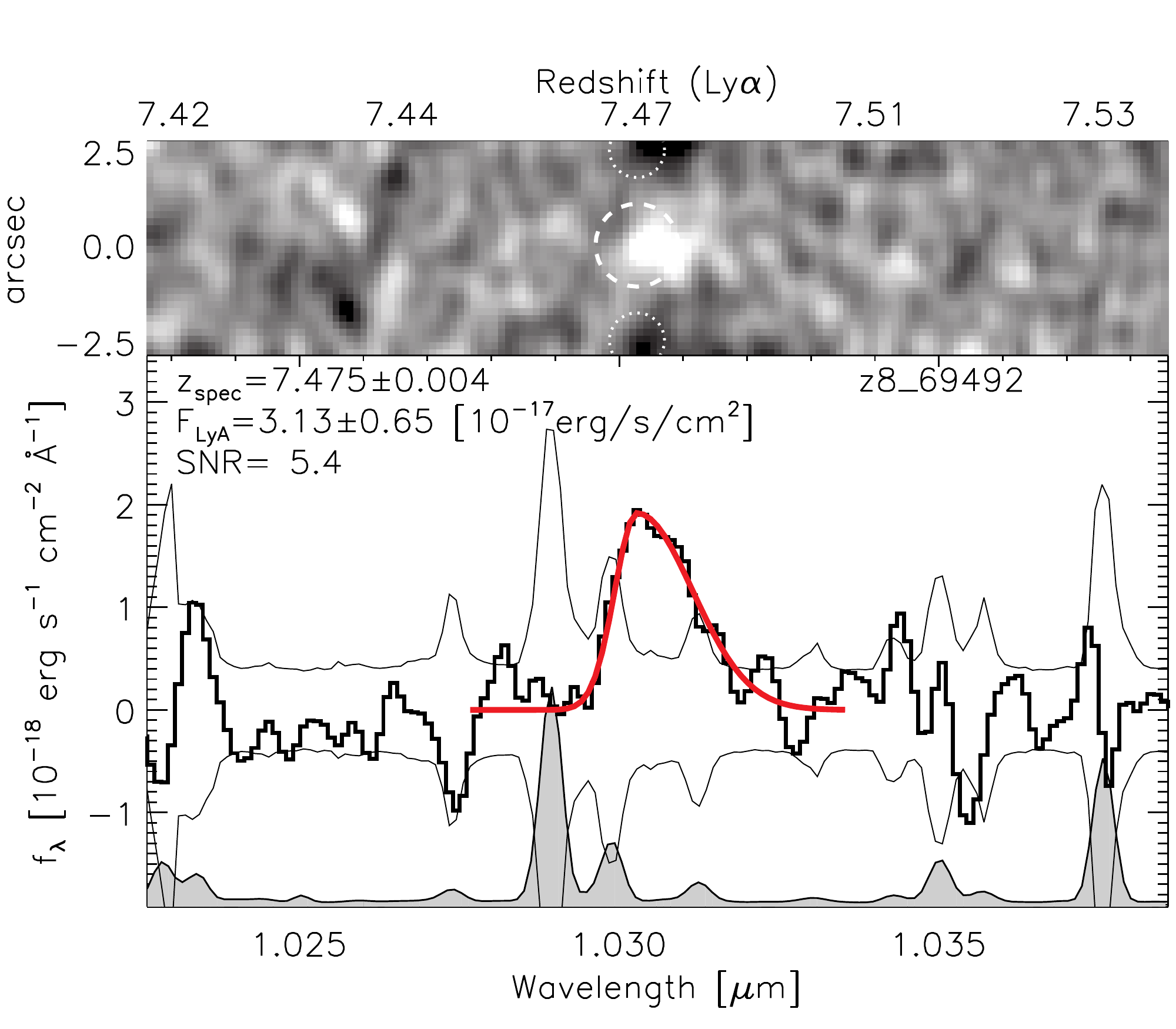}
\caption{MOSFIRE $Y$-band spectra of the detected Ly$\alpha$ emission lines for z8\_32350 (top) and  z8\_69492 (bottom). We show 2D spectra on top and 1D at the bottom. The 1D and 2D spectra are centered at the detected Ly$\alpha$ emission lines. In the 1D spectrum, the 1D signals are presented as the solid histogram, and 1$\sigma$ upper and lower bounds are shown as thin grey curves.  The red curves in 1D represent the best-fit asymmetric Gaussian curves.  The shaded curves at the bottom show sky emission areas.  Dashed circles in 2D denote the emission lines, and the negative traces caused by a dither pattern are marked with dotted circles.
}
\label{fig:mosfire}
\end{figure}

Figure \ref{fig:mosfire} shows the MOSFIRE Ly$\alpha$ spectra of z8\_32350 (top) and z8\_69492 (bottom).  The fiducial values and the errors are taken from the best-fit asymmetric Gaussian fitting to the reduced 1D spectra and 1000 resamplings of 1D spectra by perturbing 1D spectra with corresponding error spectra. The velocity offsets of Ly$\alpha$ are measured compared to the systemic redshifts estimated from the NIRSpec spectra. For z8\_13573, we take the key measurements available in \cite{Jung2022b}. For z8\_32350, \cite{Jung2022b} reported the detection of Ly$\alpha$ at $z_{\text{Ly}\alpha}=7.7482$. However, it does not match the expected wavelengths of any possible emission lines from a $z=7.5457$ system, which we measure from the [\ion{O}{iii}] and H$\beta$ emission lines from the NIRSpec observations. This object has a nearby companion which is a high-redshift candidate as well (z8\_32349; $z_{\text{phot}}\sim7.6$), marked with the cyan arrow in the HST cutout in Figure \ref{fig:689}. The potential companion is located very close to z8\_32350 in the spatial direction of the MOSFIRE slit (the yellow lines). It is therefore possible that the emission line reported in \cite{Jung2022b} is associated with the other companion object while additional observations are required to confirm its nature.  Instead, we notice a marginal detection (at a $\sim$\,2$\sigma$ level) of an emission line at $z_{\text{Ly}\alpha}=7.552$ from the MOSFIRE spectrum (Figure \ref{fig:mosfire}). The emission line is noticeable with a visual inspection, but the detection significance is low as it is found right next to a sky line. Therefore, it has been removed from the emission-line candidates in \cite{Jung2022b}.  In addition to the marginal $\sim$\,2$\sigma$-level detection, a separate MOSFIRE $Y$-band program detects the emission line at $\sim4\sigma$ (O. Cooper et al. in preparation). Thus, we consider it as a reliable detection of Ly$\alpha$. The Ly$\alpha$ properties in Table \ref{tab:elines} are calculated based on the $\sim$\,2$\sigma$-level detection spectrum. More accurate properties of Ly$\alpha$ for z8\_32350 will be further discussed in O. Cooper et al. (in preparation).

The measured Ly$\alpha$ line flux of z8\_69492 from our MOSFIRE observations, $f_{\text{Ly}\alpha}=3.13(\pm0.65)$ $\times 10^{-17}$\,erg\,s$^{-1}$\,cm$^{-2}$, is greater than the previously reported line fluxes by $\gtrsim$\,2$\sigma$ uncertainties: $1.6(\pm0.3)$ $\times10^{-17}$\,erg\,s$^{-1}$\,cm$^{-2}$ in \cite{Roberts-Borsani2016a} and $0.74(\pm0.10)$ $\times10^{-17}$\,erg\,s$^{-1}$\,cm$^{-2}$ in \cite{Stark2017a}.  Such a large variation of the measured line fluxes in MOSFIRE observations was noticed already in the two previous studies, \cite{Roberts-Borsani2016a} and \cite{Stark2017a}, and it could represent actual differences of spatially-extended Ly$\alpha$ emission \citep{Wisotzki2016a, Leclercq2017a, Song2020a, Kusakabe2022a, Bunker2023a} depending on slit position in observations \citep{Smith2019a}. The Ly$\alpha$ line flux from NIRSpec observations is also significantly smaller than what we measure from the MOSFIRE observations ($f_{\text{Ly$\alpha$,NIRSpec}}/f_{\text{Ly$\alpha$,MOSFIRE}}\sim0.2$), and this also may be partially due to significant slit loss of extended Ly$\alpha$ in NIRSpec observations. Additionally, such differences could become more significant in the high spatial resolution of NIRSpec observations while we treat our sources as point-like, but then some of the sources are spatially resolved with the NIRspec point-spread functions. Similarly, \cite{Larson2023a} find a significantly smaller Ly$\alpha$ flux in NIRSpec ($\sim$\,7\,$\times$ fainter) compared to the MOSFIRE observation. We will further discuss more detailed comparison between NIRSpec and MOSFIRE Ly$\alpha$ spectra in Section 4.2.

\subsection{Galaxy Properties from SED fitting}
We obtain the stellar population properties from galaxy spectral energy distribution (SED) fitting presented in \cite{Jung2022b}, which use the CANDELS EGS photometric catalog from \cite{Finkelstein2022a}. Briefly, the photometric catalog includes six bands of HST ACS and WFC3 broadband photometry ($V_{606}$, $I_{814}$, $Y_{105}$, $J_{125}$, $JH_{140}$ and $H_{160}$) as well as Spitzer/IRAC 3.6$\mu$m and 4.5$\mu$m band photometry.  Unfortunately, these LAE sources do not have NIRCam observations being outside the CEERS NIRCam fields.  First, galaxy SED models are constructed based on the \cite{Bruzual2003a} stellar population synthesis model with the \cite{Salpeter1955a} initial mass function and the \cite{Calzetti2001a} dust attenuation description. Then, nebular emission lines are added to model SEDs based on the \cite{Inoue2011a} emission-line ratio, and the IGM attenuation is applied according to \cite{Madau1995a}. Finally, galaxy physical properties such as stellar mass, absolute UV magnitude, dust attenuation, and star formation rate, are obtained from fitting the broadband photometry of HST and Spitzer to the model SEDs. More detailed description of SED fitting is provided in \cite{Jung2022b}. The SED-derived physical properties of our sources are listed in Table \ref{tab:elines}. 

\begin{deluxetable*}{lccc}
\tabletypesize{\footnotesize}
\tablecaption{Summary of the Measurements of three LAE targets  \label{tab:elines}} 
\tablehead{
\colhead{Source ID} & \colhead{z8\_13573} &\colhead{z8\_32350} & \colhead{z8\_69492}
}
\startdata
{MPT ID} 			& {686} & {689} & {698}\\
{R.A. (J2000.0)} 	& {215.15088} & {214.99903} & {215.05033}\\
{Decl. (J2000.0)} 	& {52.98957} & {52.94197} & {53.00745}\\
\hline
\multicolumn{4}{c}{\textbf{SED-fitting-derived Properties}} \\
{$M_{*}$ [$10^{10}$ M$_\odot$]} 	& {$0.69^{+1.19}_{-0.54}$} 	& {$0.59^{+1.30}_{-0.47}$} 	& {$0.21^{+0.40}_{-0.02}$}\\
{$M_{\text{UV}}$}				& {$-20.68^{+0.11}_{-0.11}$} 	& {$-21.98^{+0.22}_{-0.03}$} 	& {$-21.97^{+0.11}_{-0.02}$}\\
{SFR$_{\text{UV,corr}}$} 			& {57$^{+213}_{-47}$} 		& {$139^{+278}_{-107}$} 		& {$132^{+51}_{-33}$}\\
{$E(B-V)_{\text{SED}}$} 			& {$0.19^{+0.19}_{-0.19}$} 	& {$0.16^{+0.14}_{-0.16}$} 	& {$0.16^{+0.03}_{-0.03}$}\\
{$\beta$} 						& {$-1.80^{+0.47}_{-0.53}$} 	& {$-2.11^{+0.55}_{-0.57}$} 	& {$-2.05^{+0.19}_{-0.32}$}\\
\hline
\multicolumn{4}{c}{\textbf{Emission-Line Properties}} \\
{$z_{\text{sys}}$}											& {$7.7528\pm0011$} 	& {$7.5457\pm0.0001$} 		& {$7.4710\pm0.0000$}\\
{FWHM [km s$^{-1}$]$^{a}$}									& {-} 					& {$<$\,$276$} 				& {$233\pm12$} \\
{\textbf{O3} (=[\ion{O}{iii}]5007/H$\beta$)} 						& {$4.82^{+0.57}_{-0.50}$}& {$5.72^{+0.50}_{-0.42}$}	& {$6.95^{+0.41}_{-0.38}$}\\
{\textbf{O32} (=([\ion{O}{iii}]4959+[\ion{O}{iii}]5007)/[\ion{O}{ii}]3727,29)} 	& {-} 					& {$6.84^{+0.89}_{-0.64}$} 	& {$19.30^{+6.06}_{-3.40}$}\\
{\textbf{RO3} (=[\ion{O}{iii}]4363/([\ion{O}{iii}]4959+[\ion{O}{iii}]5007))} 	& {-} 					& {$0.04\pm0.01$} 			& {$0.02\pm0.00$}\\
{\textbf{R23} (=([\ion{O}{ii}]3727,29+[\ion{O}{iii}]4959+[\ion{O}{iii}]5007)/H$\beta$)}	& {-} 			& {$8.56^{+0.74}_{-0.62}$} 	& {$9.74^{+0.60}_{-0.54}$}\\
{\textbf{Ne3O2} (=\ion{Ne}{iii}3870/[\ion{O}{ii}]3727,29)} 				& {-} 					& {$0.49^{+0.09}_{-0.08}$} 	& {$1.03^{+0.35}_{-0.24}$}\\
{[\ion{O}{iii}]5007/[\ion{O}{iii}]4959} 								& {$3.26^{+0.72}_{-0.44}$} & {$2.98^{+0.18}_{-0.17}$} 	& {$2.91^{+0.10}_{-0.10}$}\\
{H$\gamma$/H$\beta$} 										& {-}					& {$0.41^{+0.08}_{-0.07}$} 	& {$0.45\pm0.05$}\\
{$E(B-V)_{\text{neb}}$$^{b}$} 									& {-} 					& {$0.28^{+0.37}_{-0.28}$} 			& {$0.11^{+0.26}_{-0.11}$}\\
{log($\xi_{\text{ion}}$/Hz\,erg$^{-1}$)$^{c}$} 					& {$25.82^{+0.39}_{-0.37}$}& {$25.24^{+0.33}_{-0.32}$} 	& {$25.44^{+0.16}_{-0.22}$}\\
{log($T_{e}\text{[\ion{O}{iii}]}$/K)} 								& {-} 					& {$4.47^{+0.07}_{-0.06}$} 	& {$4.22^{+0.05}_{-0.06}$}\\
{12+log(O/H)} 												& {-} 					& {$7.50^{+0.09}_{-0.09}$} 	& {$7.77^{+0.01}_{-0.01}$}\\
{$Z/Z_{\odot}$} 											& {-} 					& {$0.065^{+0.015}_{-0.012}$} 	& {$0.120^{+0.003}_{-0.003}$}\\
{log$U$$^{d}$} 											& {-} 					& {$-2.15^{+0.13}_{-0.12}$} 	& {$-1.76^{+0.22}_{-0.19}$}\\
\hline
\multicolumn{4}{c}{\textbf{Ly$\alpha$-Emission Properties}} \\
{$z_{\text{Ly}\alpha,\text{NIRSpec}}$} 										& {-} 					& {-} 			& {$7.483\pm0.005$}\\
{$\Delta_{\text{Ly}\alpha,\text{NIRSpec}}$ [km s$^{-1}$]} 				& {-} 					& {-} 						& {$424^{+104}_{-290}$}\\
{f$_{\text{Ly}\alpha,\text{NIRSpec}}$ [$10^{-17}$\,erg\,s$^{-1}$\,cm$^{-2}$]} 			& {-} 			& {-} 			& {$0.57\pm0.13$}\\
{$z_{\text{Ly}\alpha,\text{MOSFIRE}}$} 										& {$7.748\pm0.001$} 					& {$7.552\pm0.003$} 			& {$7.475\pm0.004$}\\
{$\Delta_{\text{Ly}\alpha,\text{MOSFIRE}}$ [km s$^{-1}$]} 			& {$-165\pm72$$^{e}$} 					& {$221\pm109$} 					& {$142\pm142$}\\
{f$_{\text{Ly}\alpha,\text{MOSFIRE}}$ [$10^{-17}$\,erg\,s$^{-1}$\,cm$^{-2}$]} 			& {$1.23\pm0.18$} 			& {$1.40\pm0.57$} 			& {$3.13\pm0.65$}\\
{EW$_{\text{Ly}\alpha,\text{MOSFIRE}}$ [\AA]} 			& {$69.1^{+29.8}_{-19.9}$} 			& {$24.0^{+18.5}_{-12.2}$} 			& {$48.7^{+15.0}_{-12.9}$}\\
{f$_{\text{esc,Ly}\alpha,\text{MOSFIRE}}$$^{f}$} 			& {$0.30^{+0.08}_{-0.07}$} 			& {$0.31^{+0.17}_{-0.14}$} 			& {$0.48^{+0.14}_{-0.12}$}\\
\enddata
\tablenotetext{}{$^{a}$\footnotesize Corrected for an instrumental broadening.\\
$^{b}$\footnotesize Based on the Balmer decrement of H$\gamma$/H$\beta$.\\
$^{c}$\footnotesize $\xi_{\text{ion}} = L_{\text{H$\beta$}}/(c_{\text{H$\beta$}}\,L_{\text{UV}})$, where the line-emission coefficient ($c_{\text{H$\beta$}}$) is 4.86$\times$10$^{-13}$\,erg for case B recombination, assuming $f_{\text{esc}}=0$. Dust correction for H$\beta$ is done based on the SED-derived E(B-V). Also, we add a systematic error of 40\% of the H$\beta$ line flux to consider the uncertainties of flux calibration based on the JWST pipeline's path-loss correction \citep{Fujimoto2023a}.\\
$^{d}$\footnotesize Ionization parameter estimated from the relation between O32 and the ionization parameter given in \cite{Papovich2022a}. \\
$^{e}$\footnotesize Based on a comparison between the Ly$\alpha$ redshift from the MOSFIRE observations and the systemic redshift from the NIRSpec prism observations. We caution that the NIRSpec wavelength calibration uncertainty could be up to $\sim$1 pixel in spectral elements, which corresponds $\sim$$3000$\,km\,s$^{-1}$ in prism spectra. Thus, additional high spectral resolution spectra are required to confirm the derived velocity offset.\\
$^{f}$\footnotesize Ly$\alpha$ escape fraction. The intrinsic Ly$\alpha$ is estimated from H$\beta$, assuming the intrinsic emission-line ratios of Ly$\alpha$/H$\alpha$ = 8.7 and H$\alpha$/H$\beta$=2.86. The observed Ly$\alpha$ fluxes are taken from the MOSFIRE observations.}
\end{deluxetable*}

\section{Results and Discussions}
\subsection{ISM Properties}
With the detected emission lines from MR grating spectra of z8\_32350 and z8\_69492, we estimate the line ratios\footnote{The emission-line diagnotics are defined in Table \ref{tab:elines}.} (O3, O32, RO3, R23, Ne3O2, [\ion{O}{iii}]5007/[\ion{O}{iii}]4959, and H$\gamma$/H$\beta$) as proxies for the ionization state, electron temperature, metal enrichment of the ISM, and dust attenuation, respectively.  The measured line ratios are listed in Table \ref{tab:elines}.  Emission-line ratios are corrected for dust attenuation based on $E(B-V)_{\text{SED}}$ values. We separately derive $E(B-V)_{\text{neb}}$ from the Balmer decrement of H$\gamma$/H$\beta$. However, the measurement uncertainties are large, thus we do not use $E(B-V)_{\text{neb}}$ for dust correction. Instead, we assume the same dust attenuation between the nebular gas and the stellar populations. Overall, the line ratio diagnostics indicate low metallicities and high ionization parameters in the ISM of the two sources. This is in general consistent with the recent findings of reionization-era galaxies from JWST observations \citep{Trump2022a, Taylor2022a, Curti2023a, Katz2023a, Sanders2023a, Tang2023a, Bunker2023a, Saxena2023a} as well as theoretical predictions on high-redshift galaxies \citep[e.g., a high O3 predicted in][as a consequnce of both low metallicity and high ionization]{Hirschmann2022a}. We note that emission-line diagnostics of these sources are also provided in \cite{Tang2023a}, and our measurements mostly agree with their measurements within the uncertainties.

\subsubsection{Ionization State}
O32 is an ionization parameter diagnostic as it represents the relative abundance between double-ionized oxygen to singly ionized oxygen \citep{Strom2018a, Kewley2019a, Papovich2022a}. It is also sensitive to metallicity as well for metal-rich gas (12+log(O/H) $>$ 9.0) \citep{Kewley2002a}, but it is not the case in our high-redshift low-metallicity galaxies with 12+log(O/H) $<$ 8.0. Recent JWST observations find O32 values roughly from 5 to 20 from reionization-era galaxies \citep{Curti2023a, Rhoads2023a, Tang2023a, Larson2023a, Sanders2023b, Mascia2023a}, that are higher than those found in low-redshift galaxies where such high O32 ratios are mostly associated with extreme [OIII] emitters \citep{Tang2019a} and/or those with high ionization parameter and low metallicity \citep{Strom2018a, Papovich2022a, Reddy2023a}. 

We measure O32 from two sources observed in MR grating as listed in Table \ref{tab:elines}: $6.84^{+0.89}_{-0.64}$ and $19.30^{+6.06}_{-3.40}$ for z8\_32350 and z8\_69492, respectively. The measured O32 ratios from our sources are also high, comparable to the range of O32 found in recent JWST studies. Particularly, z8\_69492 shows a very high O32 ratio of $\sim$20, suggesting an extreme ionization condition in the ISM.  We derive the ionization parameters (q [cm\,s$^{-1}$]) using the relation between O32 and the ionization parameters given in \cite{Papovich2022a}. The relation is obtained by modeling emission-line properties of the CLEAR survey $1.1<z<2.3$ galaxies \citep{Simons2023a} with the MAPPINGS V photoionization models \citep{Kewley2019a}. The derived dimensionless ionization parameters ($U=q/c$) are $-2.15^{+0.13}_{-0.12}$ (z8\_32350) and $-1.76^{+0.22}_{-0.19}$ (z8\_69492), listed in Table \ref{tab:elines}. These values correspond to the high end of the ionization parameters measured from lower-redshift galaxies at $z\lesssim6$ \citep[e.g,][]{Strom2018a, Papovich2022a, Reddy2023a} and consistent with the measurements of reionization-era galaxies at $z>7$ \citep{Tang2023a}. In fact, high values of O32 can be linked with potential LyC leakers \citep{Izotov2018a, Plat2019a, Chisholm2022a, Flury2022b}, and we will explore possible extreme inflation of an ionized bubble with the extreme LyC escape fraction in Section 4.4.

From the MR grating spectra of the two sources, we measure Ne3O2 ratios as well. Ne3O2 is also an effective ionization diagnostic \citep[e.g.,][]{Trouille2011a, Zeimann2015a}.  \cite{Trump2022a} examine the SMACS ERO $z>5$ galaxies in the ``OHNO" line-ratio diagram of O3 vs. Ne3O2 \citep{Backhaus2022a}, finding that the $z>5$ galaxies present higher Ne3O2 ratios compared to the low-redshift galaxies with similar O3 ratios \citep[see Figure 4 in][]{Trump2022a}. Our LAEs are found in the same regions to the SMACS $z>5$ galaxies in the OHNO diagram, requiring higher ionization and metal-poor condition within the ISM.

\subsubsection{Electron Temperature and Metallicity}
We also detect [\ion{O}{iii}]$\lambda$4363 auroral lines in the MR grating spectra of z8\_32350 and z8\_69492. The relative populations of two different collisionally-excited levels are sensitive to the gas electron temperature ($T_{e}$), and the ratio of [\ion{O}{iii}]$\lambda$4363 to [\ion{O}{iii}]$\lambda$4959+[\ion{O}{iii}]$\lambda$5007 (RO3) can be used to calculate the electron temperature in the O$^{2+}$ zone, $T_{e}\text{[\ion{O}{iii}]}$. Also, the electron temperature can be used to estimate the $T_{e}$-based metallicities based on the empirical correlations between $T_{e}$ and metallicity \citep[e.g.,][]{Amorin2015a, Perez-Montero2021a}.
We follow \cite{Trump2022a} to derive $T_{e}$ of the ISM and the $T_{e}$-based oxygen abundance, using Eq. (4) of \cite{Nicholls2020a} and Eq. (1) of \cite{Perez-Montero2021a}.  The derived values of $T_{e}$ and the oxygen abundance are listed in Table \ref{tab:elines}. The measured oxygen abundances indicate low metallicity: $Z/Z_{\odot}\sim0.07$ and $0.12$ for z8\_32350 and z8\_69492, respectively \citep[the solar oxygen abundance is 8.69 from][]{Asplund2021a}. These two galaxies are UV luminous ($M_{\text{UV}}\sim-22$) and massive (log$M_{*}/M_{\odot}\sim9.7$ for z8\_32350 and $\sim9.3$ for z8\_69492), but showing low oxygen abundances (12+log(O/H)$<$8). They fall far below the mass-metallicity relation from low-redshift galaxies at $z\lesssim3$ \citep[e.g.,][]{Henry2021a, Sanders2021a, Papovich2022a} as seen in other $z\gtrsim8$ galaxies \citep{Fujimoto2023a}.

\begin{figure}[t]
\centering
\includegraphics[width=1.05\columnwidth]{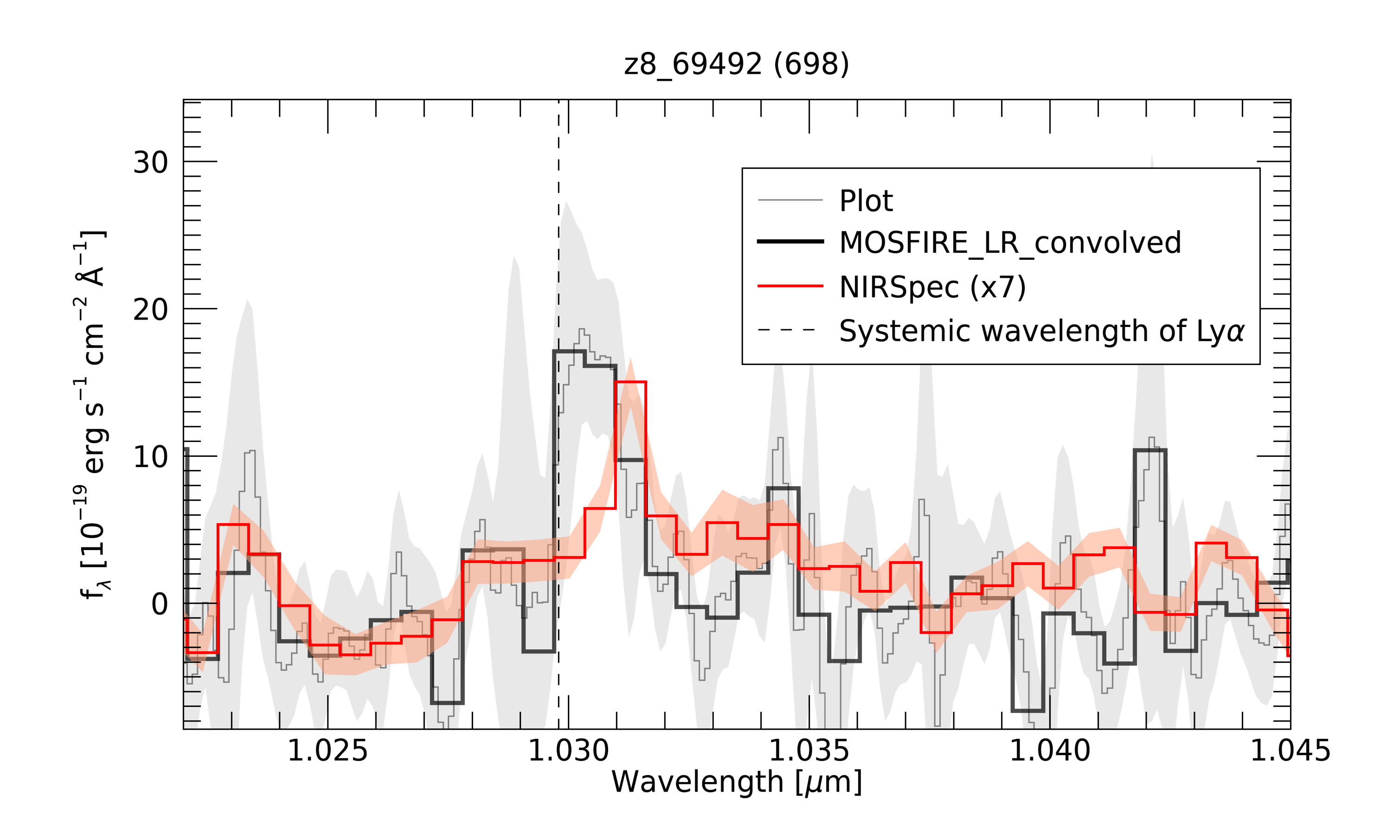}
\caption{Comparison of Ly$\alpha$ emission spectra of z8\_69492 between NIRSpec and MOSFIRE. The MOSFIRE spectrum is shown as the thin black histogram, and its degraded version to the NIRSpec MR resolution is shown as the thick black histogram. The red histogram represents the NIRSpec MR spectrum. The flux errors are shown as shaded regions. The vertical dashed line marks the systemic wavelength of Ly$\alpha$. For a visualization purpose, NIRSpec fluxes are multiplied by seven.  Notably, the Ly$\alpha$ velocity shifts are inconsistent between NIRSpec and MOSFIRE spectra.}
\label{fig:spec_comparison}
\end{figure}

\subsection{Ly$\alpha$ Velocity Offset} 
We measure the Ly$\alpha$ velocity offsets using the Ly$\alpha$ redshifts obtained from the MOSFIRE spectra (and in the NIRSpec spectra for z8\_69492), compared to the systemic redshifts obtained from the strong H$\beta$ and [\ion{O}{iii}] emission lines in the NIRSpec spectra. Due to the low spectral resolution of NIRSpec/PRISM spectra ($R\sim30$ near Ly$\alpha$), we are unable to draw a reliable measurement of a Ly$\alpha$ redshift for z8\_13573 from NIRSpec observations.  Also, no Ly$\alpha$ is seen for z8\_32350 in the NIRSpec observations.  Thus we measure the Ly$\alpha$ velocity offset from NIRSpec observations only for z8\_69492.

The derived Ly$\alpha$ velocity offsets ($\Delta_{\text{Ly}\alpha}$) are listed at the bottom in Table \ref{tab:elines}. Based on the MOSFIRE spectra, the two luminous galaxies (z8\_32350 and z8\_69492) show the Ly$\alpha$ velocity offsets of $221\pm109$\,km\,s$^{-1}$ and $142\pm142$\,km\,s$^{-1}$, respectively. These are comparable to the measurements of the Ly$\alpha$ velocity offsets in literature in a range of 110\,--\,340\,km\,s$^{-1}$ from $z>6$ galaxies with $M_{\text{UV}}\sim-22$ \citep[see Table 4 in][and the references therein]{Endsley2022b}. Notably, we find the blueshifted Ly$\alpha$ in z8\_13573. If that is true, this could be a sign of a sufficiently large ($\gg$1pMpc) ionized bubble and low residual neutral fraction in ionized regions around the galaxy \citep{Mason2020a, Park2021a, Smith2021a}. However, the Ly$\alpha$ velocity offset ($\Delta_{\text{Ly}\alpha,\text{MOSFIRE}}$) of z8\_13573 is based on a comparison of the Ly$\alpha$ redshift from MOSFIRE to the systemic redshift from the prism observations. The NIRSpec wavelength correction step may introduce the wavelength uncertainties up to $\sim$1pixel in spectral elements, which corresponds $\sim$$3000$\,km\,s$^{-1}$ in prism spectra. Thus, additional high spectral resolution observations are necessary to confirm the derived velocity offset.

With the Ly$\alpha$ spectra for z8\_69492 of both NIRSpec and MOSFIRE MR observations, we notice disagreement in the Ly$\alpha$ spectra between NIRSpec and MOSFIRE.  We present the NIRSpec and MOSFIRE spectra of z8\_69492 together in Figure \ref{fig:spec_comparison} for a detailed comparison.  
The MOSFIRE 1D fluxes are shown as the thin black histogram. For a fair comparison to the NIRSpec spectrum (red), we degrade the MOSFIRE spectra to match the NIRSpec spectral resolution, shown as the thick black histogram. The vertical dashed line represents the systemic wavelength of Ly$\alpha$. The main discrepancies are in the total line fluxes and the line profiles. We see a $\sim$\,5\,$\times$ greater flux in MOSFIRE (see Table \ref{tab:elines}). The discrepancy in the total line fluxes may indicate significant slit-loss in NIRSpec observations that is not well corrected with a point-like source assumption for this galaxy. We also find that the NIRSpec spectrum miss Ly$\alpha$ near the systemic wavelength. It is clear that the peak of Ly$\alpha$ is more redshifted from the systemic wavelength in NIRSpec than in MOSFIRE. As the NIRSpec slitlet covers a partial area of the central region in the rest-UV image (Figure \ref{fig:698}), it may lose the Ly$\alpha$ from its outskirt, resulting in the discrepancies that we discussed. Such difference in the line profile may indicate that a spatially-varying Ly$\alpha$, which experienced complex Ly$\alpha$ radiative processes in ISM and circumgalactic medium (CGM) environments \citep{Verhamme2018a, Kakiichi2018a, Song2020a, Park2022a}. 

However, as discussed above the NIRSpec wavelength correction could introduce uncertainties in the wavelength calibration within 1pixel of the spectral elements, corresponding $\sim$\,6\AA\ near Ly$\alpha$ ($\sim$200\,km\,s$^{-1}$) in NIRSpec MR grating spectra. Thus, improper wavelength correction may introduce the discrepancies to some degree that we find in velocity offsets. High spectral resolution NIRSpec observations would be needed to carry out further analysis providing an accurate measurement of the Ly$\alpha$ velocity offset.

\begin{figure*}[t]
\centering
\includegraphics[width=1.0\textwidth]{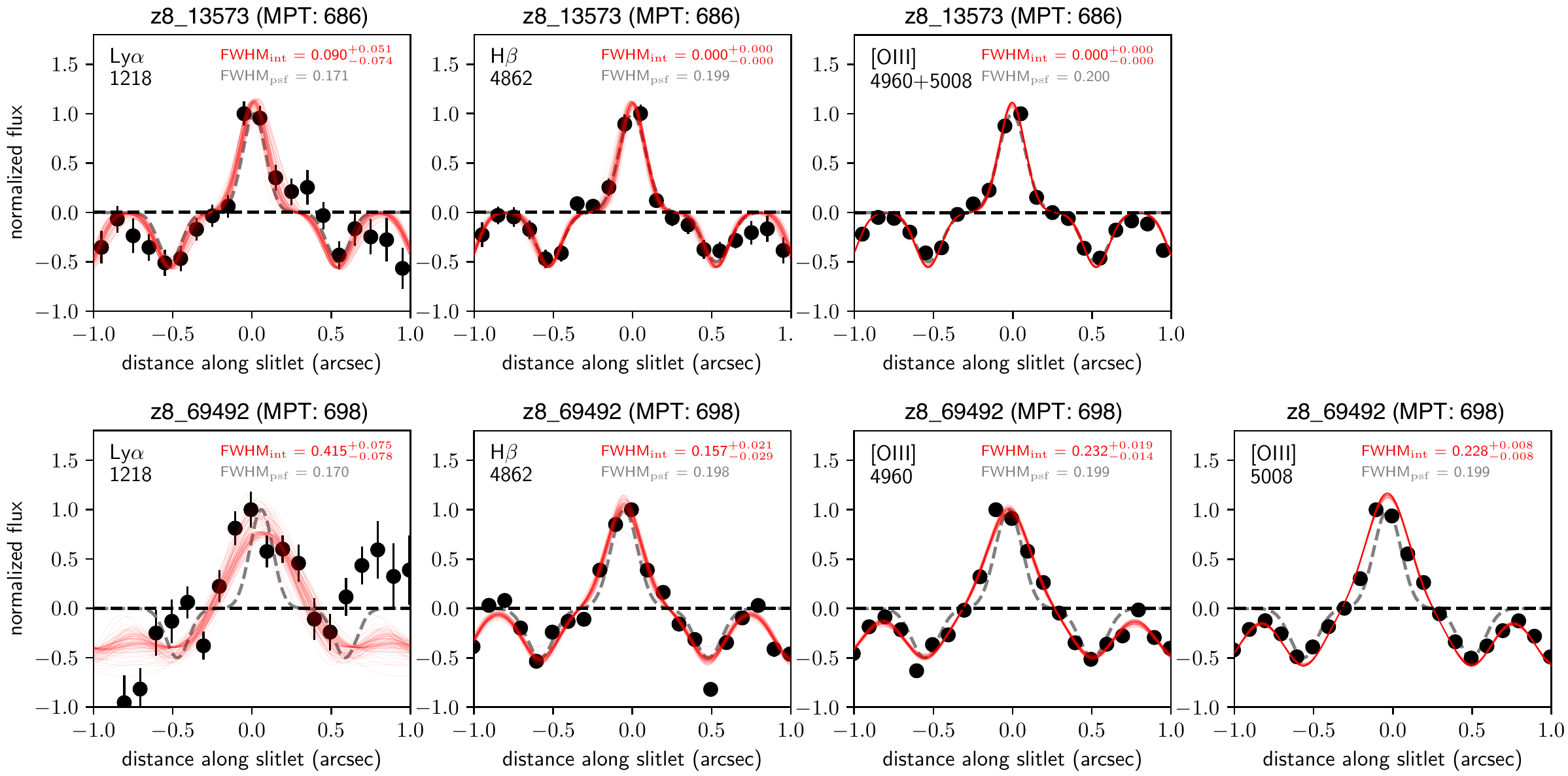}
\caption{The spatial profiles of emission lines of z8\_13573 (top) and z8\_69492 (bottom), including Ly$\alpha$ as well as H$\beta$ and [\ion{O}{iii}] lines. The dashed curves trace the spatial profiles of point-like sources while the red curves are observed spatial profiles of detected emission lines.  In each panel, the FWHM is listed in arcsec. The intrinsic FWHM values are corrected for the instrumental broadenings.  In the bottom panels, z8\_69492 features a significantly more extended Ly$\alpha$ ($2.1\pm0.4$ kpc) than its non-resonant emission lines ($\sim$1.2 kpc).  Similarly, the top panels show a tentative feature of an extended Ly$\alpha$ from z8\_13573, which is marginally resolved being larger than the size of PSF. Whereas its other emission lines are unresolved in our observations.}
\label{fig:sprofile}
\end{figure*}

\subsection{Spatially-Extended Ly$\alpha$} 
Spatially-resolved emission-line analyses of Ly$\alpha$ at $z\sim3$\,--\,$4$ often show Ly$\alpha$ halos that are spatially more extended than the UV continuum of galaxies \citep{Wisotzki2016a, Leclercq2017a, Kusakabe2022a}. A remarkable discovery of Ly$\alpha$ from a $z>$10 galaxy in \cite{Bunker2023a} also shows spatially-extended Ly$\alpha$.  Such extended Ly$\alpha$ halos may represent Ly$\alpha$ scattering through the outflowing CGM \citep[e.g.,][]{Kakiichi2018a, Leclercq2020a, Song2020a} although there are several other physical mechanisms suggested such as Ly$\alpha$ fluorescence \citep{Furlanetto2005a, Cantalupo2005a, Kollmeier2010a} and gravitational cooling \citep{Dijkstra2006a, Trebitsch2016a}.

We examine the spatial extent of emission lines of our sources whose Ly$\alpha$ seen in the NIRSpec data. Figure \ref{fig:sprofile} presents the spatial profiles of Ly$\alpha$ as well as the other prominent non-resonant emission lines (e.g., H$\beta$ and/or [\ion{O}{iii}]). In the plots, the dashed curves trace the spatial profiles of point-like sources while the red curves are observed spatial profiles of detected emission lines. Notably, z8\_69492 features a significantly more extended Ly$\alpha$ ($2.1\pm0.4$ kpc) than its non-resonant emission lines ($\sim$1.2 kpc).  We also find a tentative feature of an extended Ly$\alpha$ from z8\_13573, which is marginally resolved being larger than the size of PSF. Whereas other emission lines are unresolved in our observations. Such spatially extended Ly$\alpha$ found in our NIRSpec observations in addition to the recent discovery of \cite{Bunker2023a} may suggest a common nature of Ly$\alpha$ halos in the high-redshift universe.

\subsection{Ionized Bubbles for the Escape of Ly$\alpha$}
\subsubsection{Ionized Bubble Sizes}
In this section, we consider the possible characteristics of the ionized bubbles around these LAEs. Following \cite{Endsley2021a} and \cite{Larson2022a}, we calculate the radius of the Str\"{o}mgren spheres \citep{Cen2000a} as:

\begin{equation}
R = \left( \frac{3\,\dot{N}_{\text{ion}}\,f_{\text{esc}}\,t}{4\,\pi\,n_{\text{\ion{H}{i}}}}\right)^{1/3},
\end{equation}
where $\dot{N}_{\text{ion}}$ is the rate of intrinsic ionizing emissivity ($=L_{\text{UV}}\times\xi_{\text{ion}}$) in units of s$^{-1}$, $f_{\text{esc}}$ is the LyC photon escape fraction, and $t$ is a star formation episode time. $n_{\text{\ion{H}{i}}}$ represents the proper volume density of neutral hydrogen given as:
\begin{equation}
n_{\text{\ion{H}{i}}} = \frac{ (1-Y_{\text{He}})\,\rho_{\text{crit}}\,\Omega_{\text{b}}\,(1+z)^3}{m_{\text{H}} }
\end{equation}
where $Y_{\text{He}}$ is the primordial helium abundance by mass, $\rho_{\text{crit}}=3H^2_0/(8\pi G)$ is the critical density, $\Omega_{\text{b}}$ is the baryon density fraction.

We derive ionizing photon production efficiency ($\xi_{\text{ion}}$) from H$\beta$ emission lines following \cite{Matthee2022a}. First, we take the H$\beta$ fluxes corrected for dust attenuation and calculate $\xi_{\text{ion}} = L_{\text{H$\beta$}}/(c_{\text{H$\beta$}}\,L_{\text{UV}})/(1-f_{\text{esc}})$, where the line-emission coefficient ($c_{\text{H$\beta$}}$) is 4.86$\times$10$^{-13}$\,erg for case B recombination.   Dust attenuation for H$\beta$ is corrected based on the SED-derived $E(B-V)$.  The estimated values of $\xi_{\text{ion}}$, assuming $f_{\text{esc}}$=0, are listed in Table \ref{tab:elines}. The quoted uncertainties in the table include an additional 40\% to account for systematic errors in the flux calibration propagating into the H$\beta$ flux estimates \citep{Fujimoto2023a}. The mean value of log($\xi_{\text{ion}}$/Hz\,erg$^{-1}$) is $\sim$\,$25.5$, generally consistent with the measurements of $\xi_{\text{ion}}$ from high-redshift ($z>6$) galaxies \citep{Endsley2021a, Matthee2022a, Fujimoto2023a, Tang2023a, Bunker2023a}. 

\begin{figure*}[ht]
\centering
\includegraphics[width=0.325\textwidth]{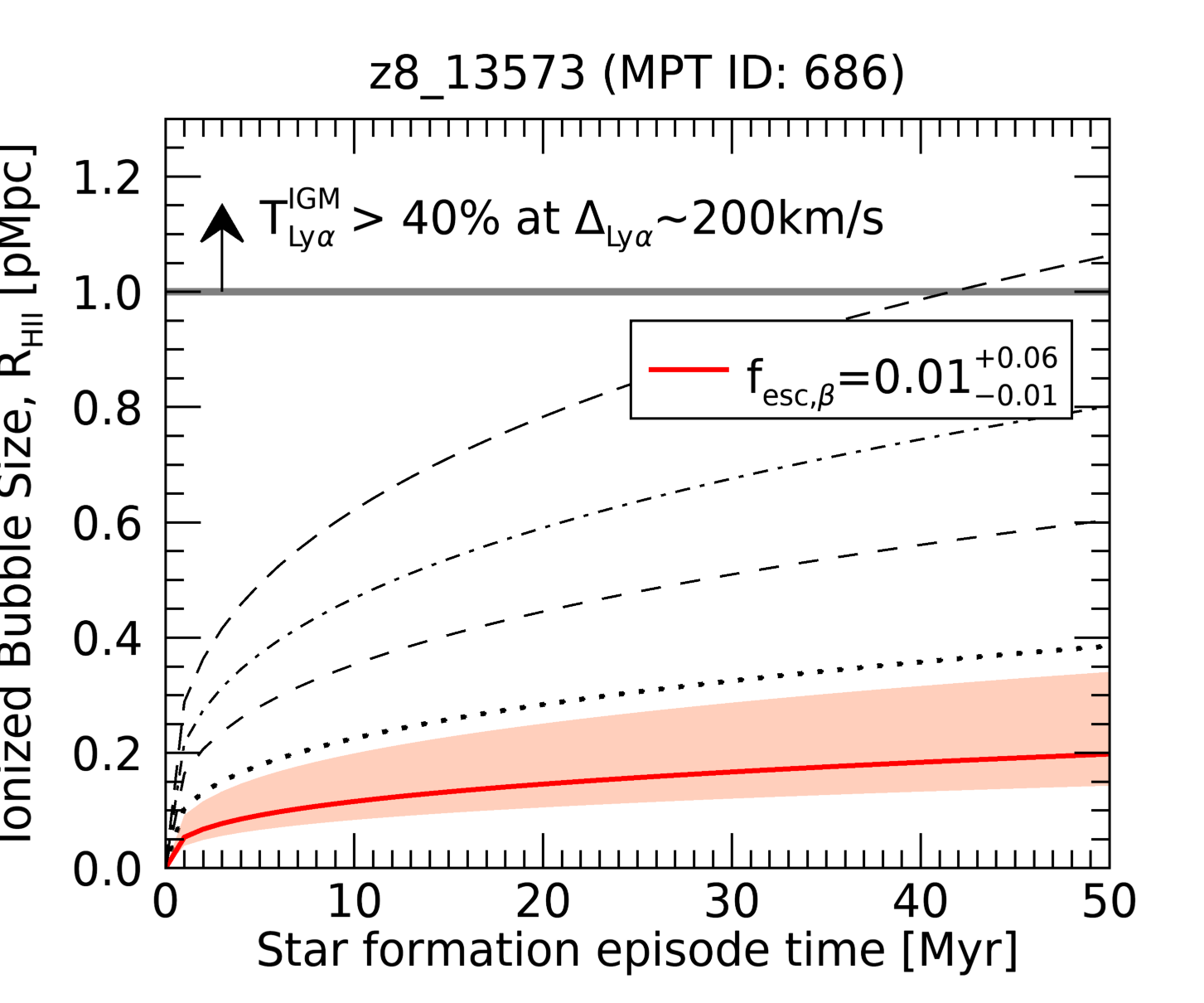}
\includegraphics[width=0.325\textwidth]{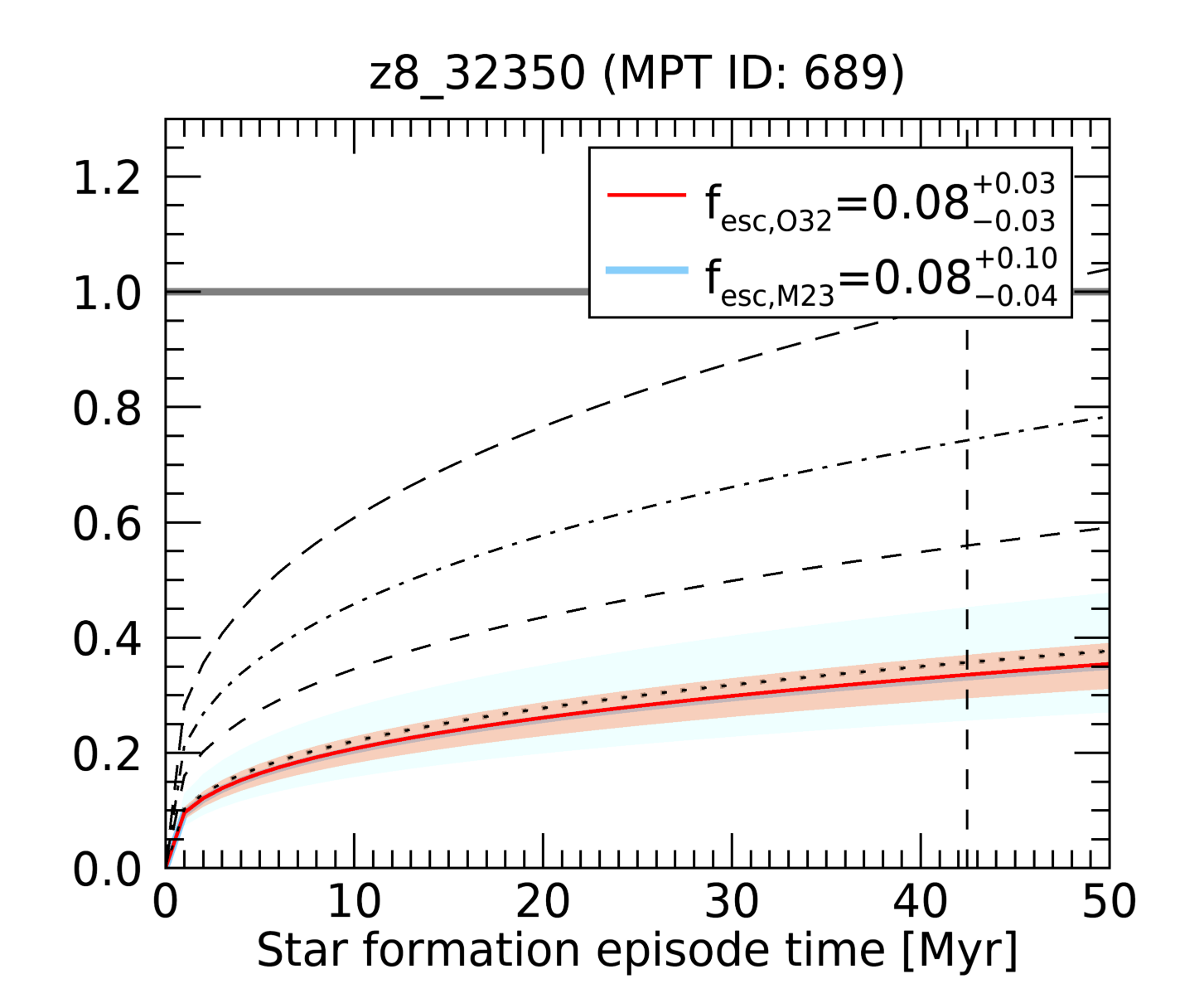}
\includegraphics[width=0.325\textwidth]{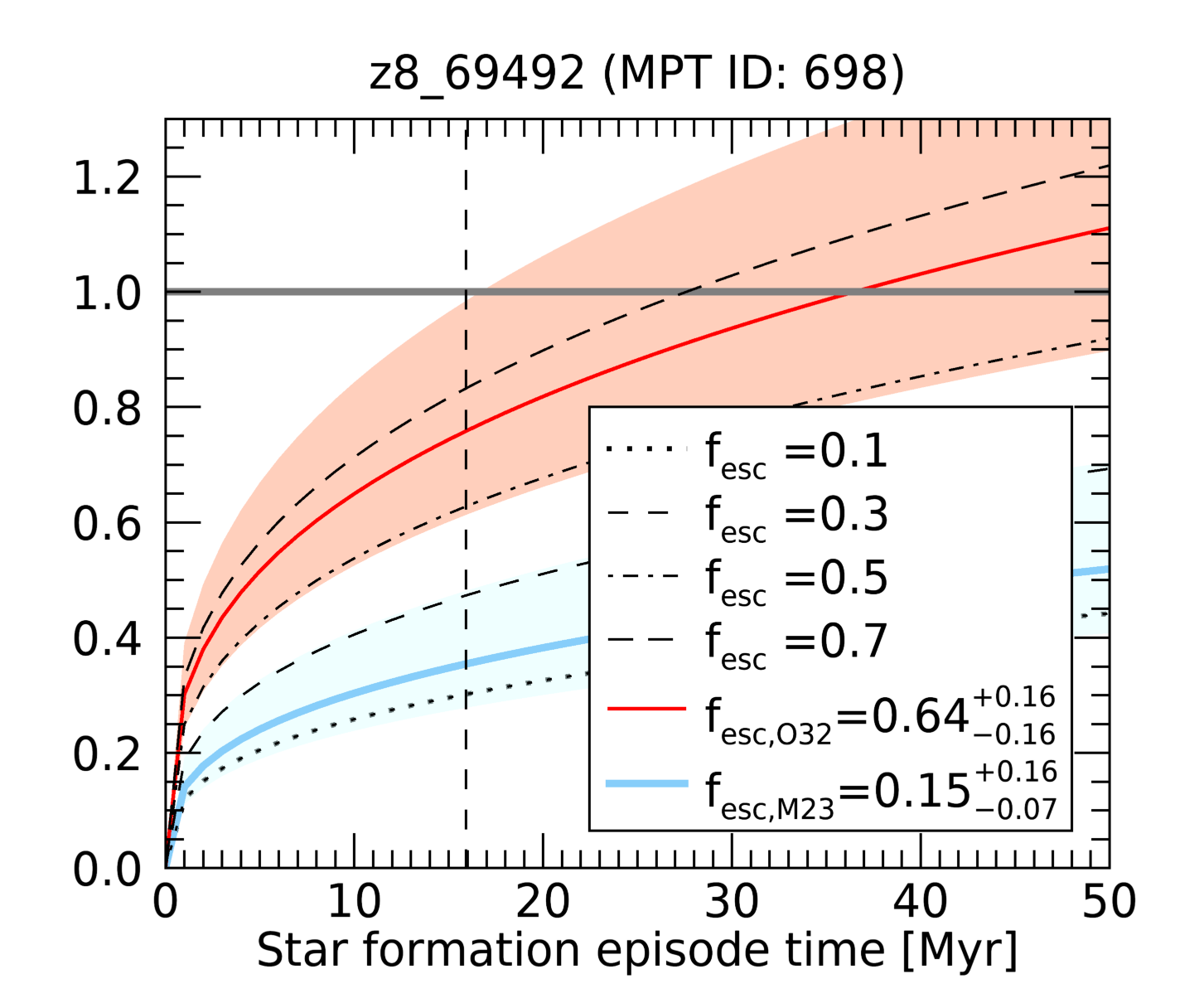}
\caption{Expected sizes of ionized bubbles around the LAEs as a function of star formation episode time for z8\_13573 (left), z8\_32350 (middle), and z8\_69492 (right). In each panel, different lines show the size growths of ionized bubbles at various values of $f_{\text{esc}}=$ [0.1, 0.3, 0.5, 0.7]. The red curves present the predictions with $f_{\text{esc}}$ values that are derived from the indirect indicators of $f_{\text{esc}}$ with the rest-UV slope \citep[$\beta$;][]{Chisholm2022a} and the O32 ratio \citep{Chisholm2018a}, and the blue curves show the predictions with $f_{\text{esc}}$ values derived from the relation given in \cite{Mascia2023a}. The gray horizontal line on top in each panel marks a characteristic bubble size for escape of Ly$\alpha$ ($R_{\text{HII}}$\,$=$\,1pMpc). The vertical dashed lines denote the maximum allowable star formation episode periods for reaching the current stellar masses of these galaxies under the assumptions of constant star formation histories. The limit on the star formation episode time for z8\_13573 is longer than 50Myr.}
\label{fig:rhii}
\end{figure*}

It is challenging to gauge the actual supplies of ionizing (or LyC) photons from galaxies mainly due to the unknown LyC escape fraction ($f_{\text{esc}}$). We calculate the predicted sizes of ionized bubbles around the three sources as a function of star formation episode time at different values of $f_{\text{esc}}=$ [0.1, 0.3, 0.5, 0.7] as shown in Figure \ref{fig:rhii}. The gray horizontal line on top in each panel marks a characteristic bubble size for escape of Ly$\alpha$ ($R_{\text{HII}}$\,$=$\,1pMpc) where Ly$\alpha$ transmission in the IGM ($T^{\text{IGM}}_{\text{Ly$\alpha$}}$) could be $\gtrsim$40\% for redshifted Ly$\alpha$ of $\Delta_{\text{Ly}\alpha}\gtrsim200$km\,s$^{-1}$ \citep[refer to Figure 1 in ][]{Mason2020a}.
The ionized bubbles mostly could not grow to $R_{\text{\ion{H}{ii}}}\gtrsim1$pMpc in a reasonable time scale ($<$50Myr) -- if powered by the observed sources alone -- except for the extreme case of $f_{\text{esc}}\gtrsim0.7$.  We calculate the maximum allowable star formation episode periods (denoted as the vertical dashed lines in Figure \ref{fig:rhii}) for reaching the current stellar masses of these galaxies by assuming constant star formation histories. That limits the growths of ionized bubbles within even shorter timescales for z8\_32350 and z8\_69492. The limit on the star formation episode time for z8\_13573 is longer than 50Myr ($\sim$120Myr).

Although z8\_32350 and z8\_69492 are very bright in UV with $M_{\text{UV}}\sim-22$, our predicted sizes of ionized bubbles suggest that they are not capable of forming ionized regions large enough for the escape of Ly$\alpha$ even with a moderately-high LyC escape of $f_{\text{esc}}=0.5$ within the allowed time scales of star formation episodes. Thus, the escape of Ly$\alpha$ at this redshift may require additional ionizing sources to create such large bubbles via local overdensities of galaxies \citep[e.g.,][]{Castellano2018a, Tilvi2020a, Endsley2021a, Larson2022a, Jung2022b, Tacchella2023a, Saxena2023a} or an extremely high $f_{\text{esc}}$ in their ISM. Without an external ionized source or an high-$f_{\text{esc}}$ ISM condition, Ly$\alpha$ could be heavily suppressed.

\subsubsection{Channels for Escape of Ly$\alpha$}
Our calculations on the size growth of ionized bubbles in the previous section suggest that the detections of strong Ly$\alpha$ may requires either additional ionizing sources to expand the bubble size further or an extremely high escape fraction of LyC photons ($f_{\text{esc}}\gtrsim0.7$) in a UV luminous galaxy. In this section, we discuss the possible explanations of individual cases of our LAEs.

First, we derive predicted LyC escape fraction of our sources. While the direct measurement of the LyC escape fraction is unavailable during the reionization era due to the IGM attenuation, it can be indirectly inferred via the proxies for the LyC escape that have been explored using low-redshift LyC sources \citep[e.g.,][]{Izotov2016a, Wang2019a, Flury2022a}. Indirect indicators of LyC escape include the escape of Ly$\alpha$ \citep{Dijkstra2016a, Verhamme2017a, Izotov2020a, Pahl2021a, Kimm2022a}, [\ion{Mg}{ii}]$\lambda\lambda$2796,2803 \citep{Chisholm2020a}, the rest-UV continuum slope \citep{Chisholm2022a}, and the O32 line ratio \citep{Izotov2018a, Chisholm2018a}.
More recently, \cite{Flury2022b} perform a statistical test of various diagnostics with the LyC measurements from the Low-redshift Lyman Continuum Survey \citep[LzLCS;][]{Flury2022a}, and \cite{Mascia2023a} provide prediction on $f_{\text{esc}}$ for reionization-era galaxies from their relation (M23 relation, hereafter) based on O32, the rest-UV continuum slope, and the size of galaxy. 

Among the indirect indicators, we focus on the O32 -- $f_{\text{esc}}$ relation first, motivated by the high O32 ratios of our sources.  We derive the LyC escape fraction based on O32 ($f_{\text{esc,O32}}$) for the two NIRSpec MR targets (z8\_32350 and z8\_69492), following the O32 -- $f_{\text{esc}}$ relation provided in \cite{Chisholm2018a}. We obtain $f_{\text{esc,O32}}=0.08^{+0.03}_{-0.03}$ for z8\_32350 and $0.64^{+0.16}_{-0.16}$ for z8\_69492. The red curves in the middle and right panels in Figure \ref{fig:rhii} represent the size evolution of ionized bubbles based on $f_{\text{esc,O32}}$. 

In the right panel (z8\_69492), based on its high $f_{\text{esc,O32}}$, z8\_69492 could form a sizable ionized region ($\sim$1pMpc) within $\sim$\,30\,--\,40 Myr of star formation episode time. This suggests possible self-driven formation of an ionized bubble around this bright galaxy that allows the escape of Ly$\alpha$ through the IGM although the allowable star formation episode time limits the bubble growth quite earlier than that (by $\lesssim$20Myr). The inferred high $f_{\text{esc,O32}}$ for z8\_69492 suggests a somewhat extreme case of LyC leakage that is rarely expected from such bright galaxy. However, \cite{Flury2022b} show that galaxies with O32 $>10$ form a high $f_{\text{esc}}$ ($\gtrsim$\,0.2) group (see their Figure 7).  Additionally, a multivariate predictor of $f_{\text{esc}}$ developed via a survival analysis technique using the LzLCS galaxies suggests $f_{\text{esc}}\sim0.7$ for z8\_69492 (A. Jaskot et al. in preparation).  Thus, the high O32-inferred $f_{\text{esc}}$ of z8\_69492 can be a reliable measurement.

However, the current diagnostics of LyC escape in general show large scatters in their empirical relations and need to be further tested. Specifically, a high O32 could be necessary for a high $f_{\text{esc}}$, but not sufficient condition \citep{Flury2022b, Mascia2023a}, which drives the high-$f_{\text{esc}}$ nature of z8\_69492 still questionable.  Indeed, the rest-UV continuum slope of z8\_69492 ($\beta=-2.05^{+0.19}_{-0.32}$) suggests a much lower $f_{\text{esc},\beta}$ at $0.04^{+0.25}_{-0.03}$ based on the relation given in \cite{Chisholm2022a}.  Reconciling both predictions of $f_{\text{esc,O32}}$ and $f_{\text{esc},\beta}$, the M23 relation given in \cite{Mascia2023a} provides the LyC escape fractions of $f_{\text{esc,M23}}=0.15^{+0.16}_{-0.07}$ for z8\_69492 although the M23 relation could underestimate $f_{\text{esc}}$ at high values of $f_{\text{esc}}>0.1$. The predicted size growths of ionized bubbles based on $f_{\text{esc,M23}}$ are shown with blue curves in the figure.  With this moderately-high $f_{\text{esc,M23}}$ value, z8\_69492 may not be fully responsible for creating a $>$1pMpc-sized ionized bubble. Thus, if this is the case, the escape of Ly$\alpha$ from this source is likely to require additional sources of ionizing photons.  In fact, \cite{Leonova2022a} suggest a galaxy overdensity around z8\_69492 with four additional high-redshift candidate galaxies although they lack spectroscopic confirmations yet.

With contradictory predictions on $f_{\text{esc}}$ for z8\_69492 based on various relations of indirect $f_{\text{esc}}$ indicators, it is difficult to draw definitive conclusion on whether or not the galaxy can supply ionizing photons sufficient for creating a $>$1pMpc-sized ionized bubble. However, the metal-poor and high-ionization ISM condition in this galaxy suggests a significant LyC leakage as same as inferred from various indirect $f_{\text{esc}}$ indicators. Thus the ionizing photon contribution form this galaxy may be enough to dominate the ionizing photon budget for reionizing the IGM around the galaxy even while requiring additional ionizing photon supplies from nearby companion galaxies.

For z8\_32350, it does not show a particularly enhanced O32 ($=6.8$), suggesting a moderate ionization state comparable to normal star-forming galaxies at this redshift.  We derive $f_{\text{esc,O32}}$ at a $\lesssim$10\% level and the M23 relation provides a similar $f_{\text{esc,O32}}$ prediction. Thus, the galaxy with the inferred $f_{\text{esc}}$ is certainly not capable of creating a sizable ionized bubble alone, again requiring additional ionizing sources for escape of Ly$\alpha$ in order to form a sufficiently large ionized bubble.  z8\_32350 is found within $\sim3$pMpc from z8\_69492 \citep{Tang2023a}. However, they are unlikely to form a contagious ionized region in given relatively large separation (as discussed in Section 4.4.1), and there is no obvious galaxy overdensity around z8\_32350 known so far. In that aspect, the escape of Ly$\alpha$ from this galaxy is somewhat puzzling.  However, we recall that there is a potential close companion galaxy (Figure \ref{fig:689}) although it requires a further spectroscopic confirmation.  Recently, \cite{Witten2023a} suggest an intriguing explanation on escape of Ly$\alpha$ with frequent galaxy mergers, finding that all reionization-era-LAE sample in their study have close companions.  Thus, the Ly$\alpha$ escape from z8\_32350 could be explained in that way with an episodic increase of Ly$\alpha$ escape through an interaction with a nearby companion.  We note that the Ly$\alpha$ properties of z8\_32350 presented in this work possess large uncertainties, and a more detailed Ly$\alpha$ properties for this source will be updated in O. Cooper et al. (in preparation). 

z8\_13573 is relatively UV-faint ($M_{\text{UV}}\sim-20.7$), thus the expected supply of ionizing photons from this source could be less than those of z8\_32350 and z8\_69492. We could not test if a high escape fraction of LyC is suggested from O32 as [\ion{O}{ii}] is not covered in its prism spectra. Instead, we measure the UV continuum slope at the rest-frame 1300--1800\AA\ ($\beta^{1550\text{\AA}}_{obs}$) from the continuum spectra seen in prism observations, and derive the $\beta^{1550\text{\AA}}_{obs}$-based LyC escape fraction $f_{\text{esc},\beta}<0.06$ at its 1$\sigma$ upper limit based on the relation given in \cite{Chisholm2022a}. This is certainly insufficient to create a large ionized bubble.  However, it emits Ly$\alpha$ comparable to those of z8\_32350 and z8\_69492. One possible explanation of such high-EW Ly$\alpha$ from z8\_13573 is the overlap of multiple ionized bubbles, being situated in a galaxy overdensity. Particularly, z8\_13573 is found in a local overdensity of LAEs having multiple LAEs within $\sim$2 pMpc radius \citep{Tilvi2020a, Jung2022b}. Also, \cite{Jung2022b} discuss the enhanced IGM transmission of Ly$\alpha$ from this galaxy by a foreground luminous galaxy. Thus, this source is prone to be located in an extended ionized region that is much larger than what can be created by this single galaxy alone.

\section{Summary and Conclusion}
We present our analysis of the CEERS NIRSpec observations for three Ly$\alpha$-emitting galaxies at $z\simeq7.47-7.75$.  The NIRSpec targets were selected as spectroscopically-confirmed LAEs from ground-based observations using the MOSFIRE spectrograph on the Keck telescope.  We analyze the emission-line properties and disagnositics of the line ratios as well as the detailed properties of Ly$\alpha$ such as the velocity offset and spatial extention of Ly$\alpha$.  We also calculate the expected size growth of ionized bubbles around the LAEs based on their inferred LyC escape fractions. Our findings are summarized as follows. 
\begin{enumerate}
    \item We analyze the ISM properties based on the nebular emission-line diagnostics and find that the LAEs are metal-poor and have high ionization condition in the ISM, consistent with recent findings of JWST/NIRSpec observations of reionization-era galaxies.
    \item We notice a difference in the spectral line profiles of Ly$\alpha$ between NIRSpec and MOSFIRE observations, which is indicative of spatially-varying properties of Ly$\alpha$ due to the complex Ly$\alpha$ radiative processes in the ISM and CGM, although high spectral resolution spectra with NIRSpec are required to perform a more detailed analysis of Ly$\alpha$. 
    \item Our NIRspec observations for z8\_69492 present Ly$\alpha$ spatially more extended than non-resonant emission lines, revealing the Ly$\alpha$ halo around this galaxy.
    \item Based on the measured emission-line properties, we compute the expected size growth of self-driven ionized bubbles around the LAEs. Our calculations suggest that escape of strong Ly$\alpha$ requires additional ionizing sources in general while a UV luminous galaxy with an extremely high LyC escape fraction could inflate an ionized region to be large enough ($\gtrsim$1pMpc) for escape of Ly$\alpha$.
\end{enumerate}

In conclusion, we reveal a complex nature of Ly$\alpha$ radiative processes from a detailed comparison of Ly$\alpha$ spectra between NIRSpec and MOSFIRE observations. Also, we find a clue to a common nature of extended Ly$\alpha$ halos around these galaxies, possibly responsible for a significant slit loss of Ly$\alpha$ in NIRSpec observations. Our findings necessitate high spectral resolution observations with a meticulous treatment on spatial variations to understand the propagation of Ly$\alpha$ photons in the era of reionization. 

Our predictions on the size growth of self-driven ionized bubbles around the LAEs suggest diverse scenarios on escape of Ly$\alpha$ during the epoch of reionization. In general, it is not feasible to create a sufficiently large ionized bubble ($>$1pMpc) for allowing escape of Ly$\alpha$ even for a UV-luminous galaxy ($M_{\text{UV}}\sim-22$) with a moderate ISM condition. 
Thus, the escape of strong Ly$\alpha$ may require additional ionizing sources from a local overdensity of galaxies around LAEs. 

z8\_13573 is relatively faint in UV compared to other two LAEs, and its contribution on ionizing photons is certainly insufficient to explain its strong Ly$\alpha$. However, the known galaxy overdensity at $z\sim7.7$ \citep{Tilvi2020a, Jung2022b} that z8\_13573 is associated with could form a contagious and extended ionized region that is sufficiently large enough for the escape of Ly$\alpha$ from this galaxy. This is showcasing how Ly$\alpha$ from relatively faint galaxies can escape the IGM being associated with other luminous companion galaxies.  z8\_69492 is also suggested to be centered at a local overdensity with its fainter companions \citep{Leonova2022a}.  However, unlikely z8\_13573, z8\_69492 may dominate the ionizing photon budget as the brightest galaxy among its fellow galaxies. z8\_32350 does not appear to be capable of creating a large bubble, neither does the galaxy have a local overdensity known yet. However, it has a potential close companion galaxy in its vicinity (as marked in Figure \ref{fig:689}), thus the detection of Ly$\alpha$ from this galaxy is possibly explained with an episodic increase of Ly$\alpha$ escape through galaxy mergers \citep{Witten2023a}.

As an alternative, we highlight an outstanding scenario of a bright galaxy with the case of an extreme ISM condition of a high LyC escape fraction creating a self-driven ionized bubble large enough for escape of Ly$\alpha$. It is suggested for z8\_69492 if the galaxy has an extremely high $f_{\text{esc}}$ as inferred from its high O32 ratio.

Our findings in this detailed case study of the LAEs during the epoch of reionization call for a further statistical analysis in future for a comprehensive understanding on escape of Ly$\alpha$ and dominant sources of ionizing photons (i.e. bright vs. faint galaxies) that are responsible for creating ionized bubbles in the neutral IGM.

\begin{acknowledgments}
This work is based on observations made with the NASA/ESA/CSA James Webb Space Telescope. The data were obtained from the Mikulski Archive for Space Telescopes at the Space Telescope Science Institute, which is operated by the Association of Universities for Research in Astronomy, Inc., under NASA contract NAS 5-03127 for JWST. These observations are associated with program JWST-ERS-01345.
\end{acknowledgments}

\bibliographystyle{aasjournal}
\bibliography{references}

\end{document}